\documentclass[preprint,authoryear,12pt]{elsarticle}

\usepackage{graphics}
\usepackage{multirow}
\usepackage{amssymb}
\usepackage{lscape}
\usepackage{color}

\journal{Icarus}

\begin{document}

\begin{frontmatter}

\title{Modelling circumplanetary ejecta clouds at low altitudes: a probabilistic approach}

\author[addr1]{Apostolos A.~Christou\corref{cor1}}
\ead{aac@arm.ac.uk}
\ead{Fax: +44 2837 527174}
\address[addr1]{Armagh Observatory, College Hill,
           Armagh BT61 9DG, Northern Ireland, UK}
\cortext[cor1]{Corresponding author}

\begin{abstract}
A model is presented of a ballistic, collisionless, steady state population of ejecta launched at randomly distributed times and velocities and moving under constant gravity above the surface of an airless planetary body. Within a probabilistic framework, closed form solutions are derived for the probability density functions of the altitude distribution of particles, the distribution of their speeds in a rest frame both at the surface and at altitude and with respect to a moving platform such as an orbiting spacecraft. These expressions are validated against numerically-generated synthetic populations of ejecta under lunar surface gravity. The model is applied to the cases where the ejection speed distribution is (a) uniform (b) a power law. For the latter law, it is found that the effective scale height of the ejecta envelope directly depends on the exponent of the power law and increases with altitude. The same holds for the speed distribution of particles near the surface. Ejection model parameters can, therefore, be constrained through orbital and surface measurements. The scope of the model is then extended to include size-dependency of the ejection speed and an example worked through for a deterministic power law relation. The result suggests that the height distribution of ejecta is a sensitive proxy for this dependency. 
\end{abstract}
   
\begin{keyword}
Dust, Dynamics \sep Impacts \sep Moon
\end{keyword}

\end{frontmatter}

% \linenumbers

\section{Introduction}
The surfaces of the Moon and other airless bodies in the solar system are subject to continuous bombardment by interplanetary meteoroids \citep{Grun.et.al1985} resulting in their being immersed in continuously-replenished clouds of impact ejecta \citep{Gault.et.al1963}. Such clouds have been detected {\it in situ} around the Galilean satellites and at distances of a few tenths of satellite radii or further away \citep{Kruger.et.al2000,Kruger.et.al2003}. For the case of the Moon there has been indirect evidence for two distinct dust populations, one at altitudes $\lesssim$ 100km \citep{McCoyCriswell1974,McCoy1976} and the other at some centimetres to hundreds of m above the surface \citep{RennilsonCriswell1974,Severny.et.al1975,Berg.et.al1976}. The mobilising action of electrostatic forces on surface dust grains has been involved in both cases \citep{Criswell1973,Rhee.et.al1977,ZookMcCoy1991,Stubbs.et.al2006}. Modelling of the ensemble properties of meteoroid impact ejecta specifically at these low altitudes - $\lesssim 0.1$ radii above the surface - and their contribution to the flux of near-surface particles has received comparably little attention. Although models that are formally valid for any distance do exist \citep{Krivov.et.al2003,Sremcevic.et.al2003}, the focus has been on reproducing measurements at higher altitudes. In particular, \citet{Krivov.et.al2003} found that the radial dependence of dust density may be approximated by a power law with an exponent of  $-5/2$ at distances of several satellite radii or larger. Data from the recent {\it LADEE} mission to the Moon are expected to shed new light on the dust environment around airless bodies \citep{Elphic.et.al2011} and justifies detailed modelling of this low-altitude component of the ejecta cloud. In this paper, we model this cloud as a steady state process and show how, under certain assumptions valid at low altitude, it is possible to derive explicitly the probability distributions of many of its properties for arbitrary ejection speed laws.
  Our aim is three-fold: (i) to fill a notable gap in the literature on analytical modelling of circumplanetary ejecta clouds in the low-altitude regime, (ii) to prepare for the interpretation of new datasets from current and future missions as well as to revisit existing datasets with new modelling tools, and (iii) to describe and demonstrate a general framework within which more sophisticated models can be constructed.
    
The paper is organised as follows: In the next Section we describe the kinematics of one-dimensional ballistic particle motion following vertical ejection from the surface. The regime where a constant gravitational acceleration can be considered a realistic approximation is also quantified here. Section 3 describes the steady state cumulative distribution and probability density functions (pdfs) of particle altitude for a given value of the (vertical) ejection speed. A methodology is then presented which uses these expressions in deriving the altitude pdfs for arbitrary - but size-independent - distributions of ejection speed. Explicit formulae for the altitude pdfs both for a uniform and a power-law probability distribution of ejection velocities are derived. In Section 4 we derive pdfs for the speed at a given altitude both in a rest frame and with respect to a moving platform. In Section 5 we lift the assumption of vertical ejection and consider the motion of ejecta in three dimensions. Section 6 introduces size-dependency to the physical process of ejection. An expanded formalism is introduced to treat this class of problems and its use is demonstrated by way of an example.
The last Section summarises our findings, discusses model limitations and the scope of this approach for the interpretation of {\it in situ} dust measurements and suggests several avenues for future work.

\section{Theoretical framework and relation to previous works}
Previous published models of dust clouds are expressed in terms of physical quantities (eg.~the grain number density as a function of with altitude) and utilise the full equations of motion for Newtonian gravity. In their model, \citet{Kruger.et.al2000} assumed vertical ejection of grains; \cite{Krivov.et.al2003} lifted that assumption by assuming a distribution of ejection directions. \citet{Sremcevic.et.al2003} extended the treatment to the asymmetric distribution of ejecta surrounding a satellite in orbit around a planet under an isotropic flux of impactors. Here, we choose to investigate the properties of the cloud within a probabilistic framework in the sense that the derived probability density functions integrate to unity. Since our focus is on the near-surface regime, we assume constant gravitational acceleration $g=GM/R^{2}$ where $G$ is the Newtonian gravitational constant and $M$, $R$ are the mass and radius of the body respectively. A consequence of this choice is that the position variable in our model is the altitude $h$ above the surface rather than the distance $r$ from the body's centre of mass. For vertical ejection velocity $\rm v_{L}$ at  ejection time $t_{L}$ the kinematic equations are
\begin{equation}  \label{eq:vh}
{\rm v}={\rm v}_{L } - g (t - t_{L})   \mbox{,      }h={\rm v}_{L}( t - t_{L})  - \frac{1}{2} g {(t - t_{L})}^{2}
\end{equation}
where ${\rm v}$ and $t$ are, respectively, the in-flight velocity and time.
From these relations one finds that the maximum altitude ${\rm h}_{\rm max}$ and corresponding time $t_{\rm max}$ are
\begin{equation}  \label{eq:htmax}
h_{\rm max} =  {{\rm v}^{2}_{L}}/{2 g}  \mbox{,      }t_{\rm max}={\rm v}_{L}/g
\end{equation}
with respect to the moment of ejection.
The kinematic equation for the altitude $h$ can then be written
\begin{equation}  \label{eq:h_htmax}
h=h_{\rm max}  - \frac{1}{2} g \left(t - t_{\rm max}\right)^{2}\mbox{.}
\end{equation}

It is useful here to know the regime where the above expressions are a good approximation to the full Kepler problem i.e.~where the gravitational acceleration varies as a function of $r$. We do this below by comparing the value of the maximum altitude achieved by a particle in the two cases.
From conservation of energy, we have
\begin{equation}\label{eq:energy}
{\rm v}^{2}= \mu \left(2/r - 1/a\right)
\end{equation}

where $a$ the semimajor axis of the keplerian ellipse and $\mu=GM$.
The semimajor axis of this ellipse may be calculated by setting ${\rm v}={\rm v}_{L}$ and $r=R$.
By setting ${\rm v}=0$, we can then solve for $r_{\rm max}$, the maximum distance from the body
\begin{equation}\label{eq:rmax}
r_{\rm max}= R / \left(1 - {\rm v}^2_{L} R/2\mu \right).
\end{equation}

If $\Delta = {\rm v}^2_{L} R/2\mu \ll 1$ we can write $r_{\rm max}\simeq R(1+\Delta)$, recovering the expression for $h_{\rm max}$ (Eq.~\ref{eq:htmax} above) for $g=\mu/R^{2}$. The relative error $\Delta h_{\rm max}/h_{\rm max}$ incurred by omitting the additional terms in the series is  $O(\Delta)$. For the case of the Moon that is relevant to the following Sections, we can assume $R=1738$ km and $\mu=4908$ $\mbox{km}^{3}$ $\mbox{sec}^{-2}$. Requiring that $\Delta<0.1$ then imposes the limit ${\rm v}_{L}<0.75$ km $\mbox{sec}^{-1}$. This corresponds to a maximum altitude of $\sim 190$ km above the lunar surface, implying that for $h \lesssim 200$ km the approximate formulas (\ref{eq:vh})-(\ref{eq:h_htmax}) may be considered as a good approximation for the lunar case. Note that the 
stated limiting ejection velocity is lower than the escape velocity from the lunar surface, $\sim 2.4$ km $\mbox{sec}^{-1}$, implying that our model ignores grains not gravitationally bound to the Moon. In the works cited at the beginning of this Section it was shown \cite[cf.~Panel a, Figure 9 of][]{Kruger.et.al2000} that the contribution of these escaping grains is minimal near the surface. Therefore, they can be safely neglected here. 
A further working assumption in our model is a ``flat surface'' approximation where the volume of space above the surface increases as a linear, rather than a quadratic, function of altitude. This incurs an error of order $\left(h/R\right)^{2} \sim 10^{-2}$ for the limiting value of $h$ stated above\footnote{An alternative formulation of this assumption is that two slabs of equal thickness $\Delta h$ have equal volume. In the full problem, these slabs map into spherical shells with a volume proportional to $r=R+h$}. Finally, we assume that all grains are ejected at the same angle from the surface normal.
In this sense, the scope of our model is intermediate between those of \citet{Kruger.et.al2000} and \citet{Krivov.et.al2003} in that we consider non-vertical, as well as vertical, ejection but not {\it distributions} of ejection angles.  

\section{Distributions of particle altitude}
\label{altitude}
In notation used here and throughout the paper, the capital symbol (e.g.~$H$) denotes a {\sl random variable}, the probability distribution of which is sought, while the corresponding small case symbol ($h$) refers to the particular {\sl value} adopted for that variable.
The cumulative distribution function (cdf) of the altitude $h$ can be arrived at with the help of Eq.~\ref{eq:h_htmax}
by considering the fraction of the total time-of-flight ($2 t_{\rm max}$) spent above $h$:
\begin{equation}  \label{eq:cdf_h}
F(h) = P(H<h) = 1 - \sqrt{1 - h/h_{\rm max}}\mbox{,          }0 \leq h < h_{\rm max}
\end{equation}
with the corresponding probability density function (pdf) being obtained by differentiation:
\begin{equation}  \label{eq:pdf_h}
f(h) = P(h<H<h+dh) = 1/ \left( 2 \sqrt{h_{\rm max}\left(h_{\rm max}-{\rm h}\right)}\right)\mbox{,           }0 < h < h_{\rm max}\mbox{.}
\end{equation}
We note that the expectation of this probability density is $E[h] = \left(2/3\right) {h}_{\rm max}$.  
Formally, to derive the probability $P$ in  Eq.~\ref{eq:pdf_h} one would still need to multiply the function $f(h)$ with the appropriate differential ($dh$ in this case)
but we opted to omit this throughout the paper to keep the expressions as simple as possible (but hopefully not simpler!).

We now assume a Poissonian process of particles being ejected from the surface with a rate parameter $\lambda$.
It can be shown that, if $1/\lambda \ll t_{\rm max}$ then Eq.~\ref{eq:pdf_h} describes the steady-state altitude distribution 
of particles for that particular value of ${\rm v_{L}}$.

\begin{figure}
\vspace{-3cm}
\centering
\includegraphics[height=8cm,angle=00]{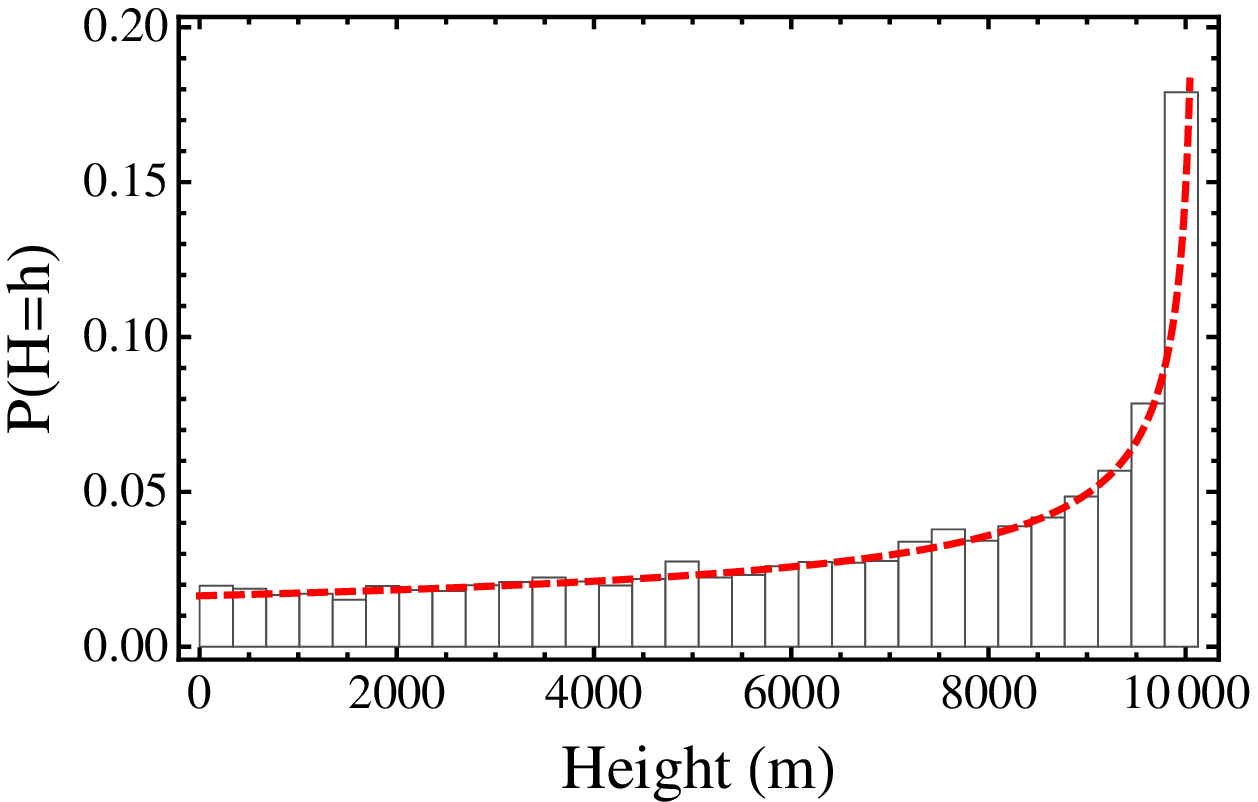}
\includegraphics[height=8cm,angle=00]{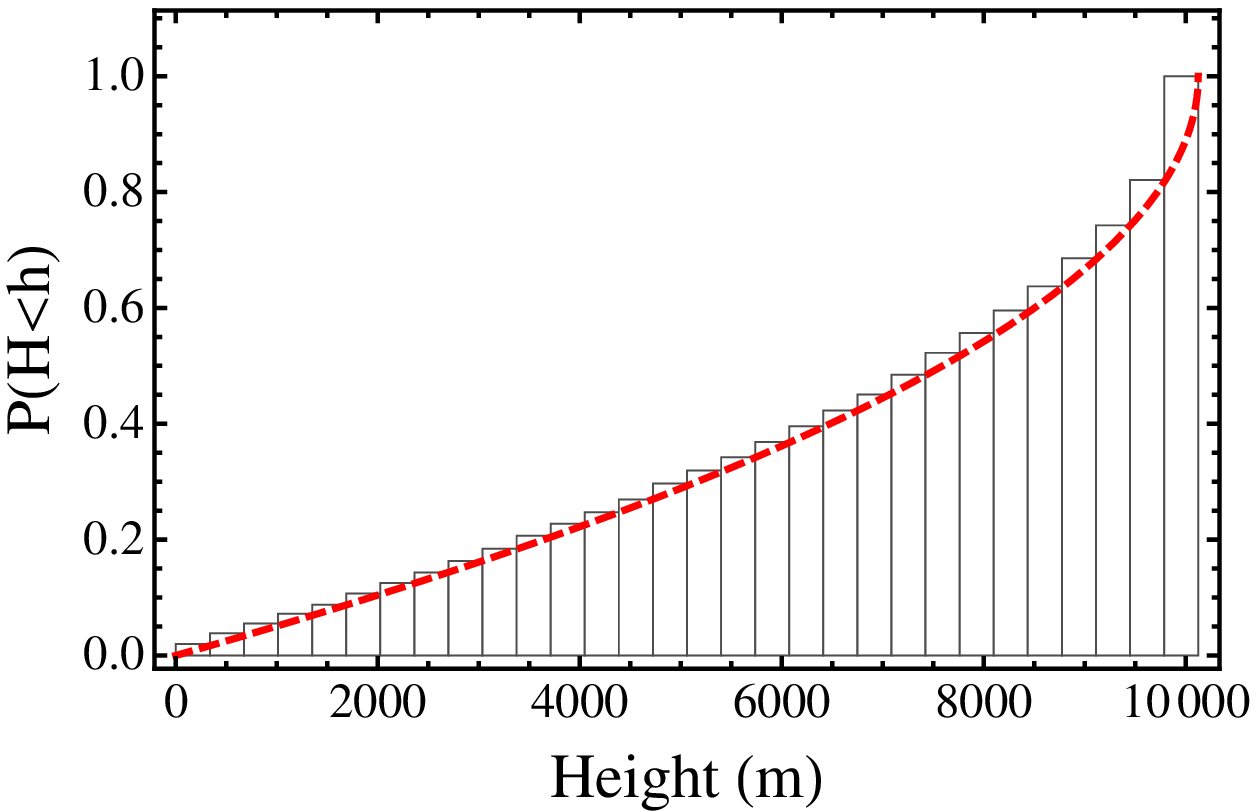}
\caption[Model (red dashed curves) and simulated (bars) distributions for the probability density function (top) and cumulative distribution function (bottom) of altitude for a population of particles launched with the same speed.]{}
\label{fig:h_distr}
\end{figure}

As an independent verification of the above expressions, 
we have generated $N=10^{4}$ uniformly distributed random variates of launch time within the 
interval $\left[0, 2 t_{\rm max}\right]$, calculated the respective altitudes using Eq.~\ref{eq:h_htmax} at  some random time 
with respect to the start of the simulation, 
plotted the resulting probability density and cumulative distributions and superposed the curves described by Eq.~\ref{eq:pdf_h} and Eq.~\ref{eq:cdf_h} respectively. 
The result is shown in Fig.~\ref{fig:h_distr} where we have set $g = 1.6$ m $\mbox{sec}^{-2}$ and 
${\rm v}_{L}=180$ m $\mbox{sec}^{-1}$. 
The effective rate parameter is $\left(1/2\right) 10^{4} / t_{\rm max} \simeq 45$ $\mbox{sec}^{-1}$ and $t_{\rm max}=112$ sec 
so the requirement for a steady state is satisfied. Note the cusp at $h=h_{\rm max}$ due to 
 the denominator of the pdf in Eq.~\ref{eq:pdf_h} and its first derivative vanishing for that value of the altitude. 

Next, we allow the ejection speed to vary according to a probability density function $g({\rm v}_{L})$. The resulting steady-state pdf for the altitude at a given time $t$ may be evaluated as: 
\begin{equation}  \label{eq:fvofh}
f_{\rm V} (h) = N(h<H<h+dh)/N(H>0)
\end{equation}

where
\begin{equation}  \label{eq:nhgt0}
N(H>0)= \int \int \lambda g({\rm v}_{L}) d \Delta t_{L} d {\rm v}_{L}
\end{equation}

and
\begin{equation} \label{eq:nofh}
 N(h<H<h+dh)=\int \int \lambda g({\rm v}_{L})  f(h) d \Delta t_{L} d {\rm v}_{L}\mbox{.}
\end{equation}

The integration over $\Delta t_{L} = t-t_{L}$ is due to the fact that only particles that are in flight (ie $\Delta t_{L} < 2 t_{\rm max}$) can contribute to the pdf.

For example, let us assume a {\it uniform} ejection speed distribution   
\begin{equation}  \label{eq:pdf_v}
g({\rm v}_{L}) = P({\rm v}_{L}<V_{L}<{\rm v}_{L}+d{\rm v}_{L}) = 1/\left({\rm v}^{+}_{L}-{\rm v}^{-}_{L}\right)\mbox{,           } {\rm v}^{-}_{L}\leq {\rm v}_{L} \leq {\rm v}^{+}_{L}\mbox{.}
\end{equation}

Eq.~\ref{eq:pdf_v} is not, strictly speaking, realistic in the sense that it is not discussed or considered 
in the literature in relation to modelling ejecta populations. It is inserted here mainly as a paedagogical tool 
to demonstrate the methodology advocated in this work and to highlight the emergence of certain features 
as one transitions from the monotachic\footnote{From the greek $\tau \alpha \chi \acute{\upsilon} \tau \eta \varsigma$ = velocity} (single value of ejection speed) regime to a {\it distribution} of ejection speeds for the particles. As shown later in the paper, such features are also observed when more physically relevant ejection speed 
models are adopted.

Upon double integration, Eq.~\ref{eq:nhgt0} yields
\begin{equation}
N(H>0)= (\lambda /g)  \left({\rm v}^{+}_{L} + {\rm v}^{-}_{L}\right)
\end{equation}

Expression \ref{eq:nofh} can be evaluated using the transformation $\cos^{2} S = 1 - 2gh/{\rm v}^{2}_{L}$.
It becomes
\begin{equation}
 N(h<H<h+dh)=2 \lambda / \left({\rm v}^{+}_{L} - {\rm v}^{-}_{L}\right) \int \sin^{-1} S d S 
\end{equation}

Substituting these into Eq.~\ref{eq:fvofh} then yields the altitude pdf
\begin{equation}\label{eq:pdf_h_for_g}
f_{\rm U}(h) =-2 g \left(B_{u}\left({\rm v}^{+}_{L}, h\right) - B_{u}\left(\max \{ {\rm v}^{-}_{L}, \sqrt{2gh}\},h\right)\right) / \left({\rm v}^{+ 2}_{L} - {\rm v}^{- 2}_{L}\right) 
\end{equation}
where
\begin{equation}\label{eq:b_explicit_g}
B_{u}\left({\rm v}_{L}, h\right)  =  \log \left( \tan \left( \arcsin \left(\sqrt{2 g h / {\rm v}^{2}_{L}}\right)/2\right)\right)
\end{equation}
and the subscript ``U'' refers to the uniform speed distribution assumed.
  
\begin{figure}
\vspace{-3cm}
\centering
\includegraphics[height=4cm,angle=00]{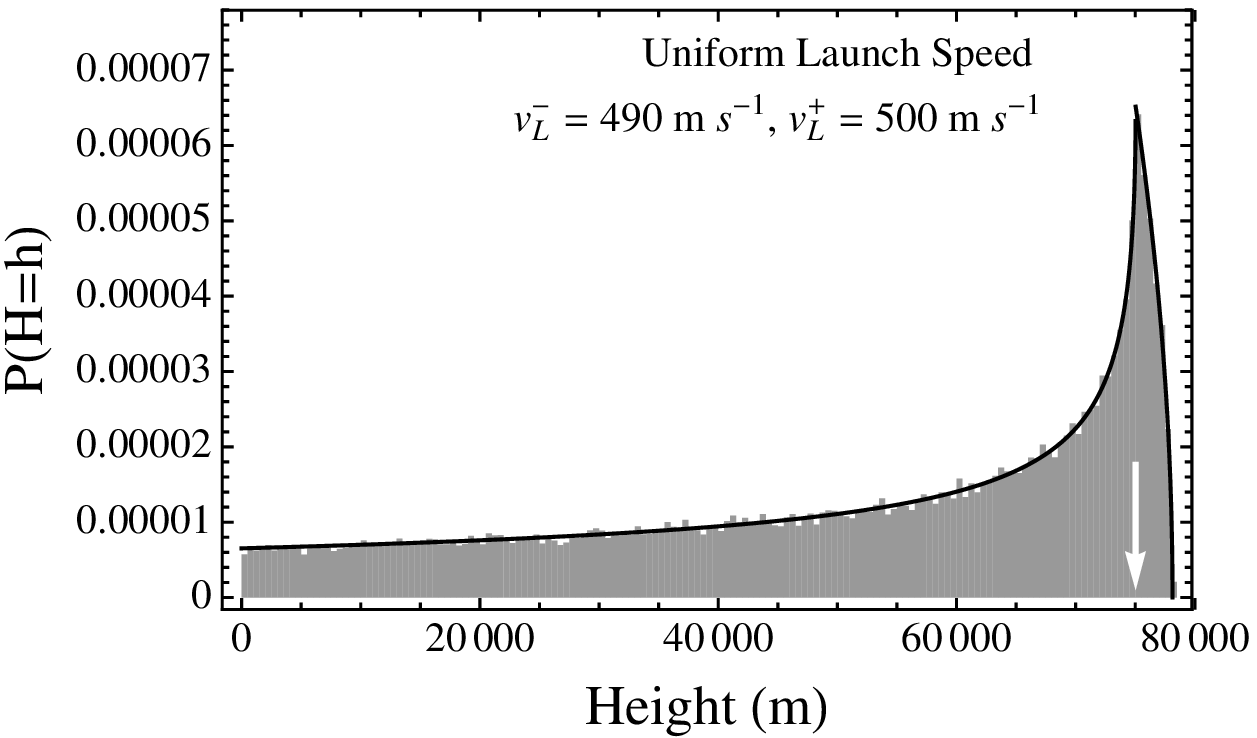}
\includegraphics[height=4cm,angle=00]{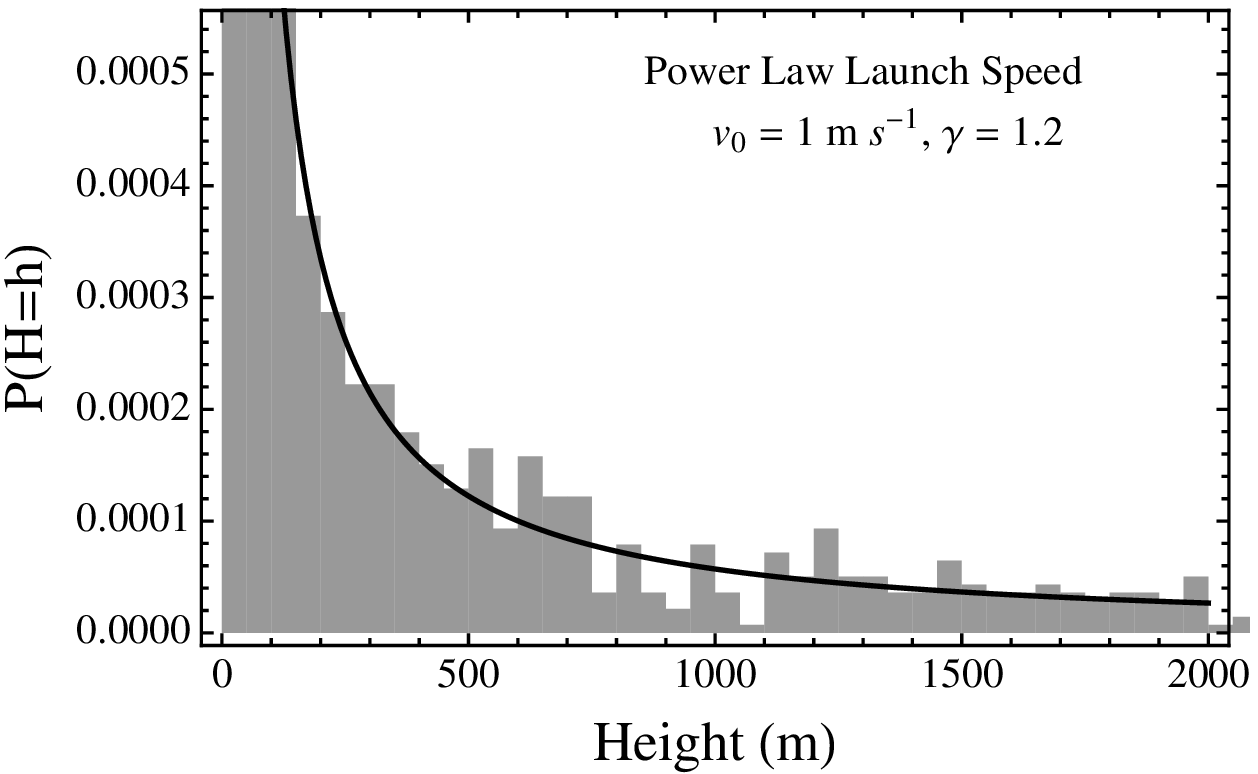}\\
\includegraphics[height=4cm,angle=00]{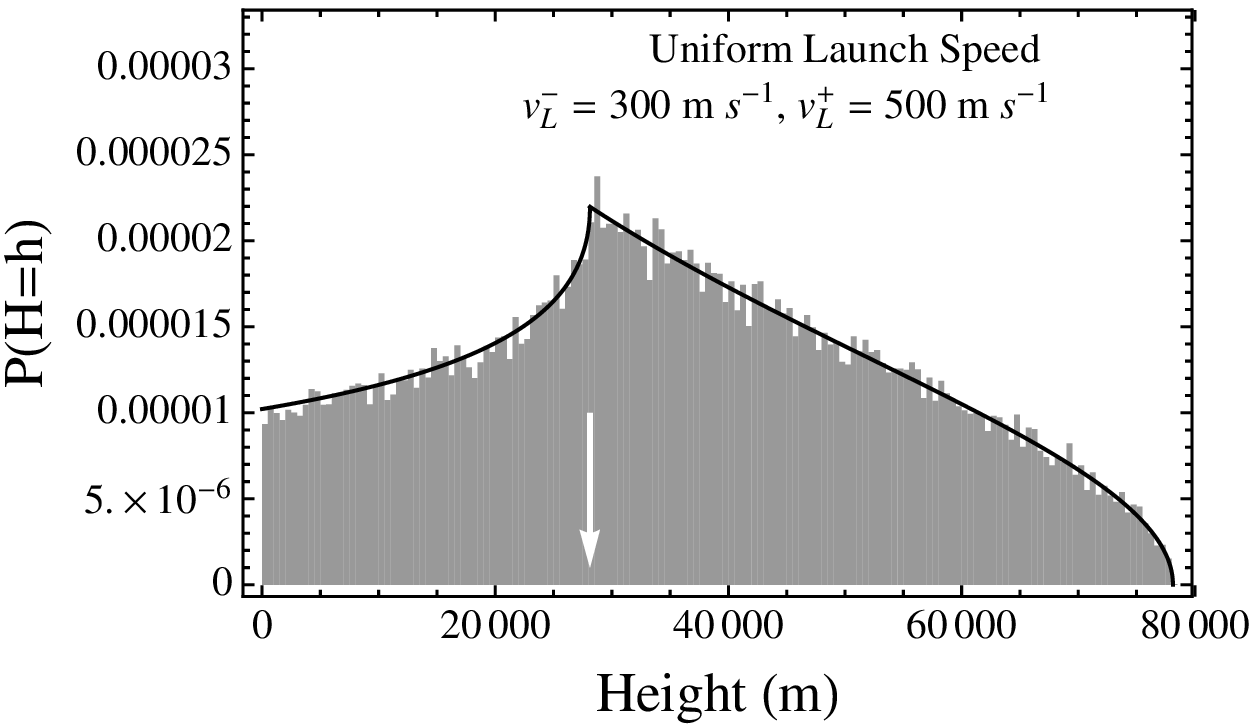}
\includegraphics[height=4cm,angle=00]{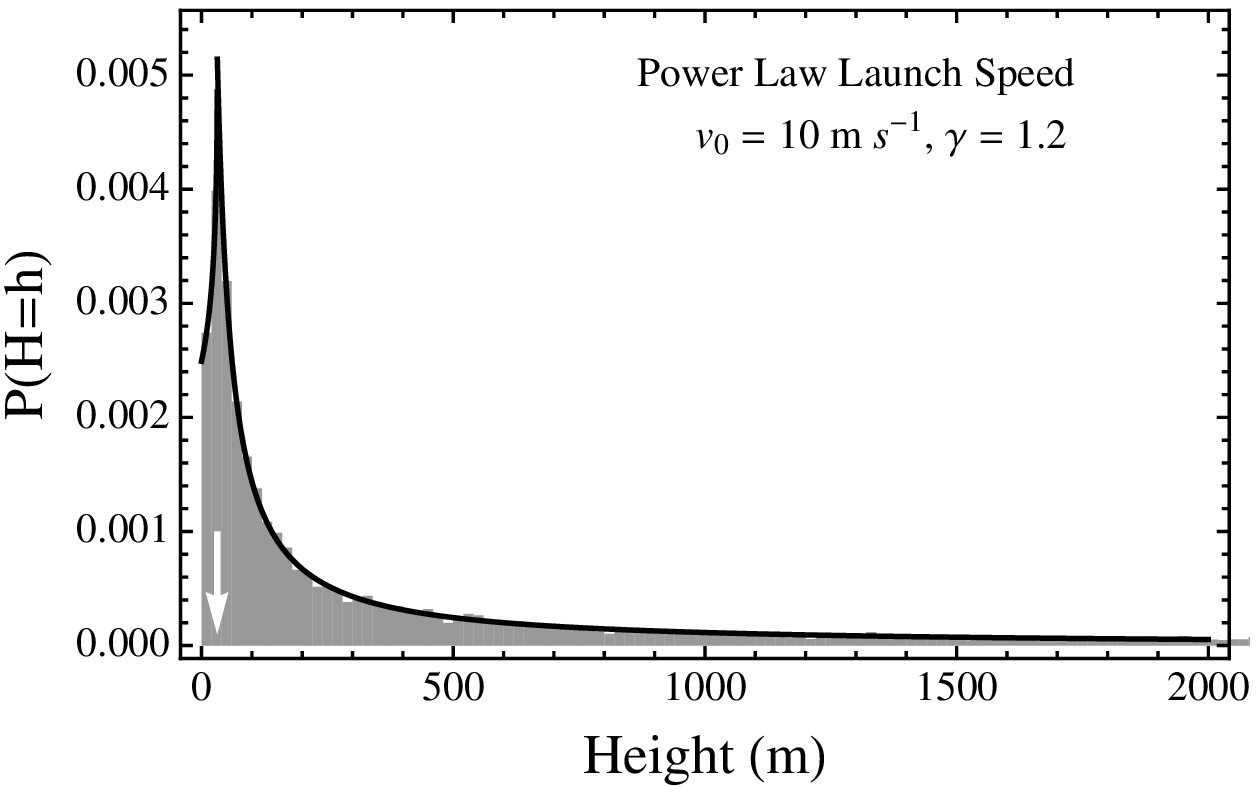}\\
\includegraphics[height=4cm,angle=00]{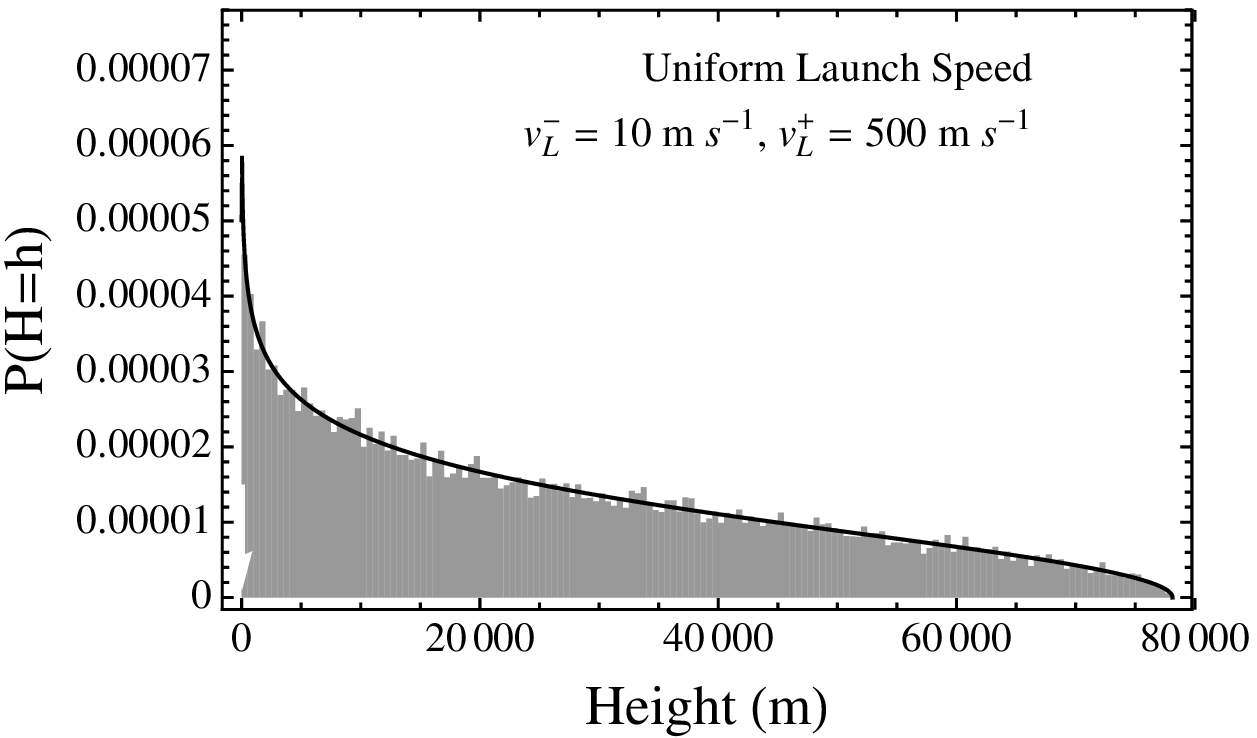}
\includegraphics[height=4cm,angle=00]{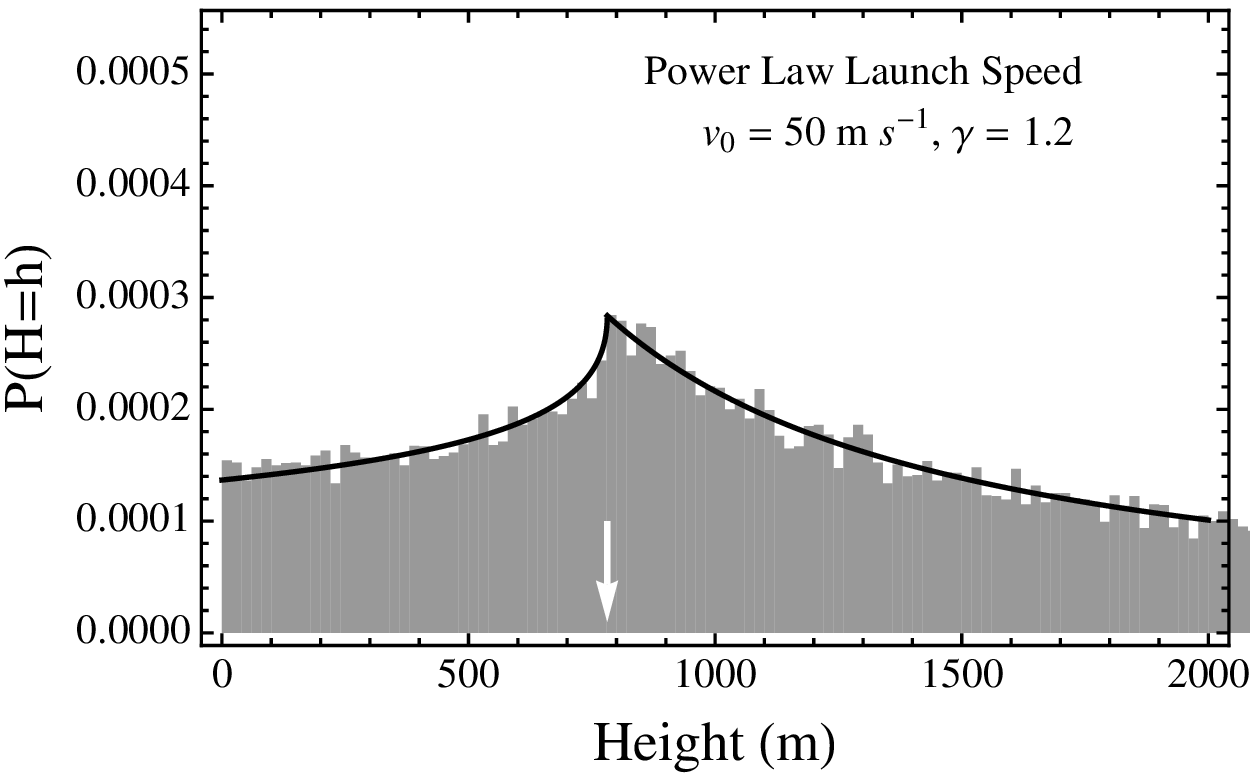}
\caption[Model (bold curves) and simulated (gray histogram) distributions of altitude for uniformly distributed (left panels) and power-law distributed (right panels) ejection speeds.]{}
\label{fig:hv_distr}
\end{figure}

To verify this result, we utilise uniform random variates of launching time as before. $N=10^{5}$ variates are generated within the fixed time interval $\left[0, 1000\right]$ sec for a rate parameter of 100 $\mbox{sec}^{-1}$. We assign to each particle a launch speed value randomly drawn from the distribution given by Eq.~\ref{eq:pdf_v} and for three different choices of the defining parameters: (a) ${\rm v}^{-}_{L}=490$ m $\mbox{sec}^{-1}$, ${\rm v}^{+}_{L}=500$ m $\mbox{sec}^{-1}$ (b) ${\rm v}^{-}_{L}=300$ m $\mbox{sec}^{-1}$, ${\rm v}^{+}_{L}=500$ m $\mbox{sec}^{-1}$, and (c) ${\rm v}^{-}_{L}=10$ m $\mbox{sec}^{-1}$, ${\rm v}^{+}_{L}=500$ m $\mbox{sec}^{-1}$.  We note for future reference that $t_{\rm max}({\rm v}^{+}_{L})=312.5$ sec while  $h_{\rm max}({\rm v}^{+}_{L})=78$ km. The results of the simulation and at $t=700$ sec are shown in the left panels of Fig.~\ref{fig:hv_distr} where we have overplotted the analytical solution (black curve).
 For case (a), the resulting distribution is similar to that for a constant ejection speed, as can be intuitively 
expected. For case (b), a maximum occurs near $h=h_{\rm max}({\rm v}^{-}_{L})$ (indicated by the white arrow). 
That, and the cusp seen in the analytical laws, are due to the fact that particles launched with ${\rm v}^{-}_{L}<{\rm v}_{L}<\sqrt{2gh}$ (Eq.~\ref{eq:htmax}) no longer contribute in Eq.~\ref{eq:pdf_h_for_g} as 
$h$ increases beyond the value $h_{\rm max}({\rm v}^{-}_{L})$. 
 
 An ejection speed law relevant to modelling ejecta production from satellite surfaces \citep{Krivov1994,Kruger.et.al2000,Krivov.et.al2003,Kruger.et.al2003} has the form
 \begin{equation} \label{eq:pdf_v_gamma}
 p({\rm v}_{L}) = \left( \gamma / {\rm v}_{0} \right) \left({\rm v}_{L}/{\rm v}_{0}\right)^{-\gamma -1}\mbox{,     }{\rm v}_{L} > {\rm v}_{0}
 \end{equation}
 where ${\rm v}_{0}$ and $\gamma$ are positive constants. 
 
 Following the same procedure as above we obtain
 \begin{equation} \label{eq:pdf_h_for_p_denom}
N(H>0)= 2 \lambda \left({\rm v}_{0}/g\right) \gamma / \left(\gamma - 1\right)
\end{equation}

and
 \begin{equation} \label{eq:pdf_h_for_p_num}
N(h<H<h+dh)= -2 \lambda \gamma {\rm v}^{\gamma}_{0} {\left(2 g h\right)}^{-(\gamma + 1)/2} \int_{S{|}_{u=1}}^{0} \sin^{\gamma} S d S 
\end{equation}
where $\cos^{2} S = 1 - 2gh/\left(u^{2}{\rm v}^{2}_{0}\right)$, $u = {\rm v}_{L}/{\rm v}_{0}$.

Substituting in Eq.~\ref{eq:fvofh} and evaluating the integral yields
\begin{equation}\label{eq:pdf_h_for_p}
\hspace{-0mm}f_{\rm P}(h) = g(\gamma \hspace{-1mm}-\hspace{-1mm}1){\left(2 g h\right)}^{-(\gamma + 1)/2}{\rm v}^{\gamma-1}_{0}\hspace{-1mm}\left[B_{p}\hspace{-0mm}\left(1\right) \hspace{-0.5mm}-\hspace{-1mm} B_{p}\hspace{-1mm}\left(\hspace{-0.5mm}\max{\hspace{-1mm}\{\sqrt{1\hspace{-1mm}-\hspace{-1mm}2gh/{\rm v}^{2}_{0}},0\}\hspace{-0.5mm}}\right)\right] 
\end{equation}
where
\begin{equation}\label{eq:b_explicit_gamma}
B_{p}\left(x \right)  = x\ast {_{2}F_{1}}\left(1/2, (1 - \gamma)/2,3/2,x^{2}\right),
\end{equation}
 
 $_{2}F_{1}(.)$ denoting the Gauss hypergeometric function and the subscript ``P'' referring to the power law assumed for the ejection speed distribution. 
  
The right panels of Fig.~\ref{fig:hv_distr} show these analytical expressions overplotted on the result of a numerical simulation where $N=2 \times 10^{6}$ variates are generated within the fixed time interval $\left[0, 4000\right]$ sec for a rate parameter of 500 $\mbox{sec}^{-1}$. The launch speed value was randomly drawn from the distribution given by Eq.~\ref{eq:pdf_v_gamma} with $\gamma=1.2$ and for three different choices of ${\rm v}_{0}$: (a) ${\rm v}_{0}=1$ m $\mbox{sec}^{-1}$, (b) ${\rm v}_{0}=10$ m $\mbox{sec}^{-1}$ and (c) ${\rm v}_{0}=50$ m $\mbox{sec}^{-1}$. The altitude statistics shown were collected  at $t=2000$ sec. To speed up the Monte Carlo simulations we imposed a cutoff value for the ejection speed of ${\rm v}_{\rm cutoff}=800$ m $\mbox{sec}^{-1}$. This changes slightly the model to the effect that the expression for $N(H>0)$ needs to be multiplied by a factor of $ 1 - {\left( {\rm v}_{\rm cutoff}/{\rm v}_{0}\right)}^{1 - \gamma}$. Formally, one also needs to change the upper limit of the integral in Eq.~\ref{eq:pdf_h_for_p_num}; however, since most of the power is in the small end of the speed spectrum, the result remains essentially unchanged.
 
\begin{figure}
\vspace{-3cm}
\centering
\includegraphics[height=8cm,angle=00]{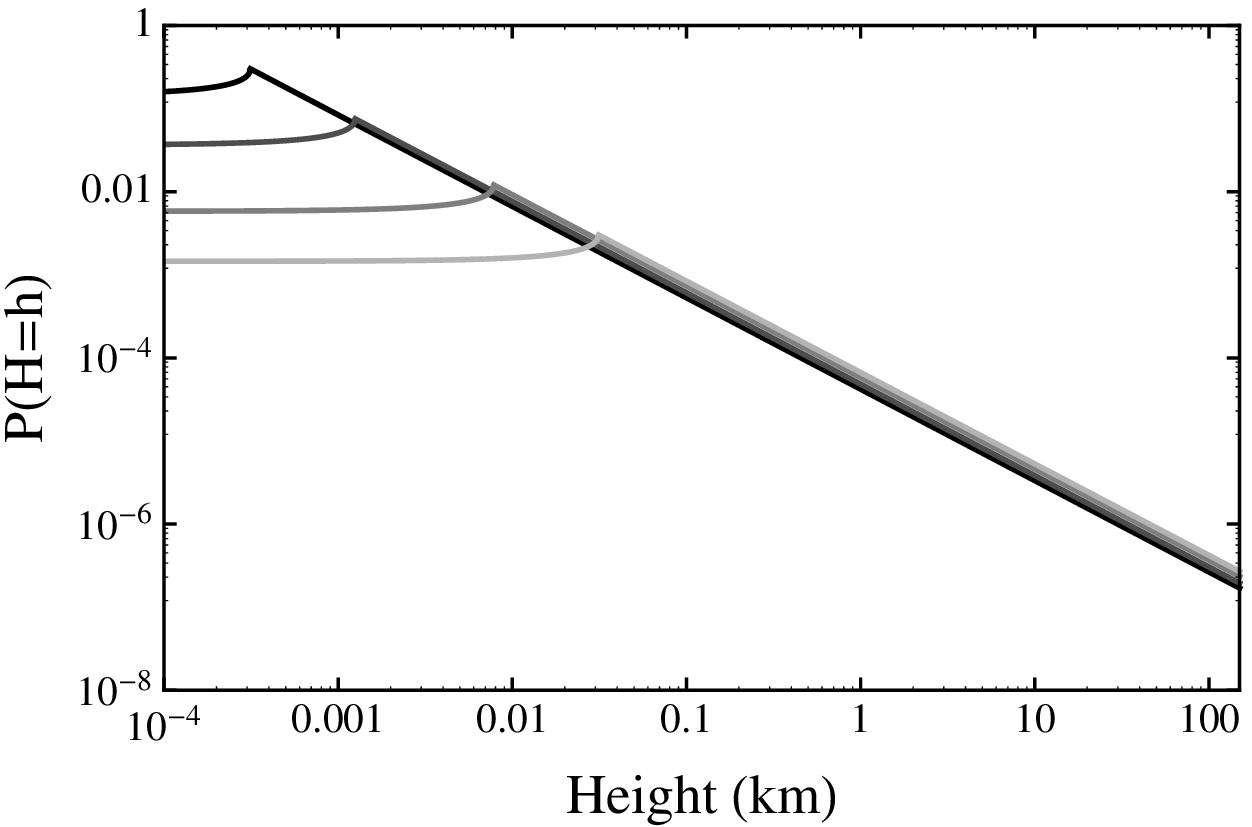}
\includegraphics[height=8cm,angle=00]{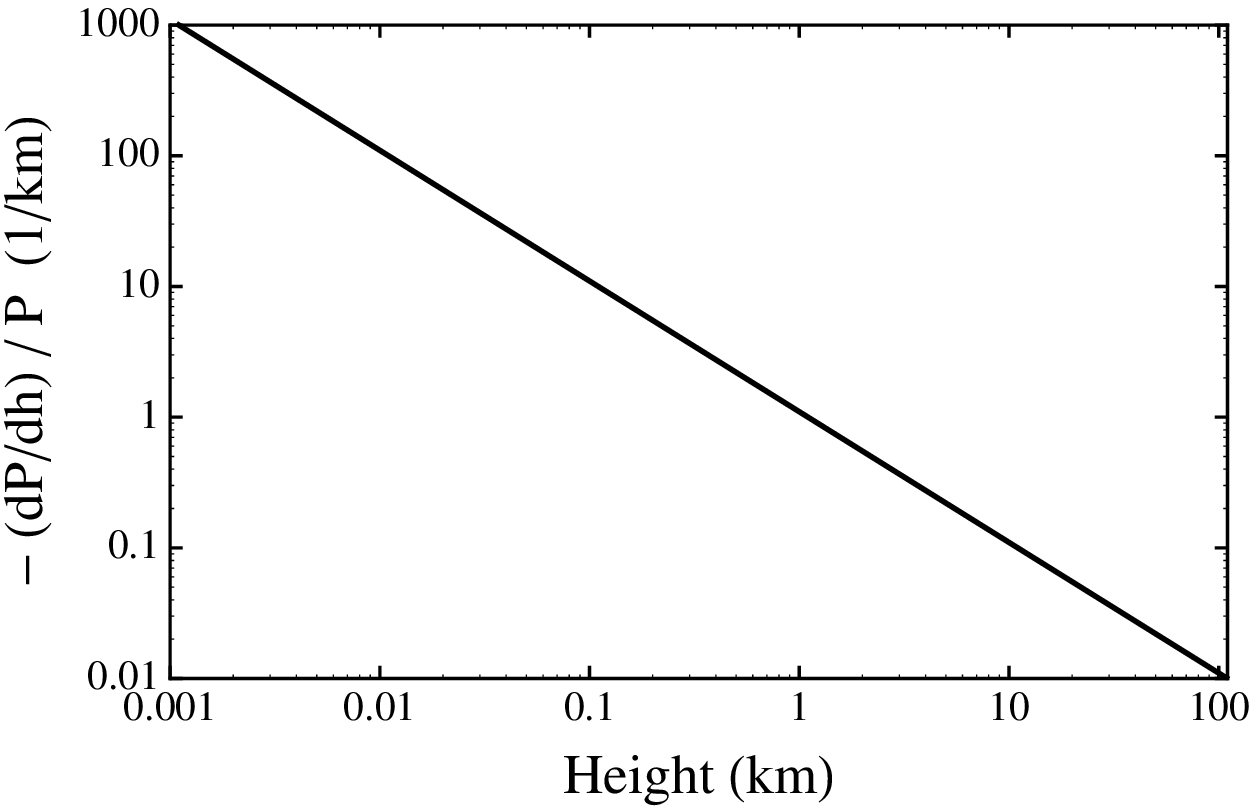}
\caption[Top: Altitude probability density functions for a power-law speed distribution and different values of the ejection speed limit ${\rm v}_{0}$ with a lighter colour corresponding to a higher value. Bottom: Effective inverse scale height for the distribution with ${\rm v}_{0} = 1$ m $\mbox{sec}^{-1}$.]{}
\label{fig:h_distr_sh}
\end{figure}

Certain properties of the altitude distribution are worthy of further comment. A cusp, seen for the case of the uniform speed distribution (left panels), is also present here and corresponds to $h=h_{\rm max}({\rm v}_{0})$ (indicated by the arrow). In the top panel of Fig.~\ref{fig:h_distr_sh}, Eq.~\ref{eq:pdf_h_for_p} is plotted for different values of ${\rm v}_{0}$: 1, 2, 5 and 10 m $\mbox{sec}^{-1}$. The scale in both axes is logarithmic. There we see that $P_{max}=P(h_{max})$ is consistently higher than the value at the surface $P_{0}(=P(h=0))$.
As ${\rm v}_{0}$ increases, these both decrease and the power under $P(h)$ for $h<h_{max}$ gradually spreads to higher altitudes. Formally, one can trace these features to the expressions for $P_{0}$ and $P_{max}$ in terms of the model parameters $g$, ${\rm v}_{0}$ and $\gamma$:
\begin{equation}
P_{0}=\frac{g}{{\rm v}^{2}_{0}} \frac{\left( \gamma - 1\right)}{\left( \gamma + 1\right)}\mbox{, } P_{max} = \sqrt{\pi}P_{0}\frac{\left(\gamma + 1\right)}{\gamma} \frac{\Gamma\left(\frac{\gamma + 1}{2}\right)}{\Gamma\left(\frac{\gamma}{2}\right)}
\end{equation}
where $\Gamma$ denotes the Gamma function. Observe that both probabilities are inversely proportional to ${\rm v}^{2}_{0}$ and that the ratio of $P_{max}/P_{0}$  depends only on the speed exponent $\gamma$ and is greater than unity \footnote{For $0<\gamma< 3$,  $P_{max}  > 0.9 \sqrt{\pi} P_{0}$}. On the far side of the cusp, the same value of $P$ as $P_{0}$ is attained at $h=\left({\rm v}^{2}_{0}/2 g \right)[\left(\sqrt{\pi}/2\right)\left(\gamma+1\right) \Gamma(\frac{1+\gamma}{2})/ \Gamma(1+\frac{\gamma}{2})]^{2/\left(\gamma+1\right)}$, bounding the near-surface region with the highest number density. 

For $h>h_{max}$ the profiles are linear in log-log space ie $\log P(h) / \log{h}$ is constant. This follows from Eq.~\ref{eq:pdf_h_for_p} where the quantity in square brackets becomes independent of $h$ so taking the $\log$ of both sides results in a linear law with slope $-(\gamma+1)/2$. A consequence of this property is that $\frac{d P}{dh}/P$ decreases with increasing altitude (bottom panel of Fig.~\ref{fig:h_distr_sh}). In other words, the ``effective scale height'' i.e.~the inverse of the plotted quantity, is altitude-dependent and increases as $h$ increases.   

As a final note in this Section, we show how the functional form of Eq.~\ref{eq:pdf_h_for_p} can be retrieved by imposing the assumptions of Section 2 on the model by \citet{Krivov.et.al2003}. In that work, particles are ejected within a cone of opening half-angle $\psi_{0}$ (their Fig.~3) and with a distribution of ejection angles given by their Eq.~21. Their results and those in this Section should agree for the case $\psi_{0}=0$ (vertical ejection of grains).   
Using their notation we find ${\tilde{u}}^{2} \simeq {\tilde{v}}^{2} + h/r_{M}$ to first order in $h/r_{M}$, then set $1 - (h/r_{m}) = \cos^{2} S$ and change the dummy variable in their Eq.~44 from $\tilde{u}$ to $S$. The $-\left(\gamma+1\right)/2$ power-law dependence on $h$ is obtained from their Eq.~42 as the zero-order term in the power-series expansion of $\tilde{r}^{-2}$ in terms of $h/r_{M}$.  
 
 \section{Distributions of particle speed at a given altitude}
 \label{speed_at_altitude}
Another important descriptor of the ejecta population is the distribution of speed ${\rm v}$ at a given altitude $h$.
The pdf of this distribution may be expressed as
\begin{equation}  \label{eq:pdf:v_given_h}
 P(V={\rm v}|H=h)= P(V={\rm v}\mbox{, }H=h)/P(H=h)
 \end{equation}	

 where $P(V={\rm v}\mbox{, }H=h)$ is the joint probability density of particles with speeds between ${\rm v}$ and  ${\rm v}+d{\rm v}$ and altitudes between $h$ and $h+dh$. The reader is reminded that, in the notation of conditional probability, the quantities on the right-hand side of the separator `` $|$ '' are treated as constant parameters while those on the left-hand side denote random variables.
 
 Considering the transformation $\left( {\rm v}_{L}\mbox{, }\Delta t_{L}=t-t_{L}\right) \rightarrow   \left( h\mbox{, }{\rm v}\right)$ we can write  
  \begin{eqnarray}\label{vh2vldtl} 
P  (V_{L}={\rm v}_{L}(h\mbox{, }{\rm v})  \mbox{, }\Delta T_{L}=\Delta t_{L}(h\mbox{, }{\rm v}))   ||J|| d{\rm v} dh  & = &  \nonumber \\
 P(V_{L}={\rm v}_{L}  \mbox{, }\Delta T_{L}=\Delta t_{L}) d {\Delta t_{L}} d {\rm v}_{L}  &  &
 \end{eqnarray}
 
where the right-hand side represents the number of particles launched between  $\Delta t_{L}$ and $\Delta t_{L}+d \Delta t_{L}$ ago and with launch speeds between ${\rm v}_{L}$ and ${\rm v}_{L}+d{\rm v}_{L}$, $J$ is the Jacobian of the transformation and $||.||$ denotes the absolute value of its determinant. From Eq.~\ref{eq:vh} we find $||J||={\rm v}^{-1}_{L} \left({\rm v}\mbox{, } h \right)$ while the joint density distribution for $\Delta t_{L}$ and ${\rm v}_{L}$ can be expressed as
 \begin{equation}  \label{eq:vl_tl}
P(V_{L}={\rm v}_{L}  \mbox{, }\Delta T_{L}=\Delta t_{L})=P(\Delta T_{L}= \Delta t_{L} | V_{L}={\rm v}_{L})P_{t}(V_{L}={\rm v}_{L})
 \end{equation} 
 
For a steady state process, the first term on the right hand side of Eq.~\ref{eq:vl_tl} is the reciprocal of the time-of-flight, $2 {\rm v}_{L}/{g}$. The second term is the probability that a particle in flight at time $t$ was launched with speed ${\rm v}_{L}$.

 For a uniform ejection speed distribution, the latter probability is proportional to the time-of-flight. The constant of proportionality may be found by integrating the probability over all possible values of  ${\rm v}_{L}$. The joint probability density of ${\rm v}_{L}$ and $\Delta t_{L}$ is then
 \begin{equation}  \label{eq:vl}
 P(V_{L}={\rm v}_{L}  \mbox{, }\Delta T_{L}=\Delta t_{L})=g/ \left({\rm v}^{+ 2} - {\rm v}^{- 2}\right)
 \end{equation} 
 and from Eq.~\ref{vh2vldtl} we find
 \begin{equation}  \label{eq:jpd_vh}
  P(V={\rm v}\mbox{, }H=h)= g / \left(\left({\rm v}^{+ 2} - {\rm v}^{- 2}\right) \sqrt{{\rm v}^{2} + 2gh} \right)
   \end{equation} 

Making use of Eq.~\ref{eq:pdf_h_for_g} in Section \ref{altitude}, Eq.~\ref{eq:pdf:v_given_h} then becomes
\begin{eqnarray}  \label{eq:v_given_h_uni}
P(V={\rm v}|H=h) &=&- (1/\sqrt{{\rm v}^{2}+2 g h})/ \nonumber \\
        &   &\left(B_{u}\left({\rm v}^{+}, h\right) - B_{u}\left(\max \{ {\rm v}^{-}, \sqrt{2gh}\},h\right)\right)\mbox{, }\nonumber \\
    &   &  \sqrt{{\rm v}^{- 2} - 2 g h} < {\rm v} <  \sqrt{{\rm v}^{+ 2} - 2 g h}
 \end{eqnarray}
 where the quantity $B_{u}$ is from Eq.~\ref{eq:b_explicit_g}. 
 
 For the power law distribution given by Eq.~\ref{eq:pdf_v_gamma}, proceeding along similar lines yields
 \begin{eqnarray}  \label{eq:v_given_h_gamma}
P(V={\rm v}|H=h) &=& \frac{{\left(2 g h\right)}^{(\gamma + 1)/2}} {{\left({\rm v}^{2}+2 g h\right)}^{(\gamma + 2)/2}} /  \nonumber \\
 &   & \left[B_{p}\left(1\right) - B_{p}\left(\sqrt{\max\{1 - 2gh/{\rm v}^{2}_{0},0\}}\right)\right]\mbox{, }  \nonumber \\
   &   &  {\rm v} > \sqrt{{\rm v}^{2}_{0} - 2 g h} 
 \end{eqnarray}
 
  where the quantity $B_{p}$ is now that of Eq.~\ref{eq:b_explicit_gamma}. For this case, the second term in Eq.~\ref{eq:vl_tl} is of the form
  $P_{t} \propto{\rm v}_{L} \times {\rm v}^{-\gamma-1}_{L}$.
 
\begin{figure}
\vspace{-3cm}
\centering
\includegraphics[height=4cm,angle=00]{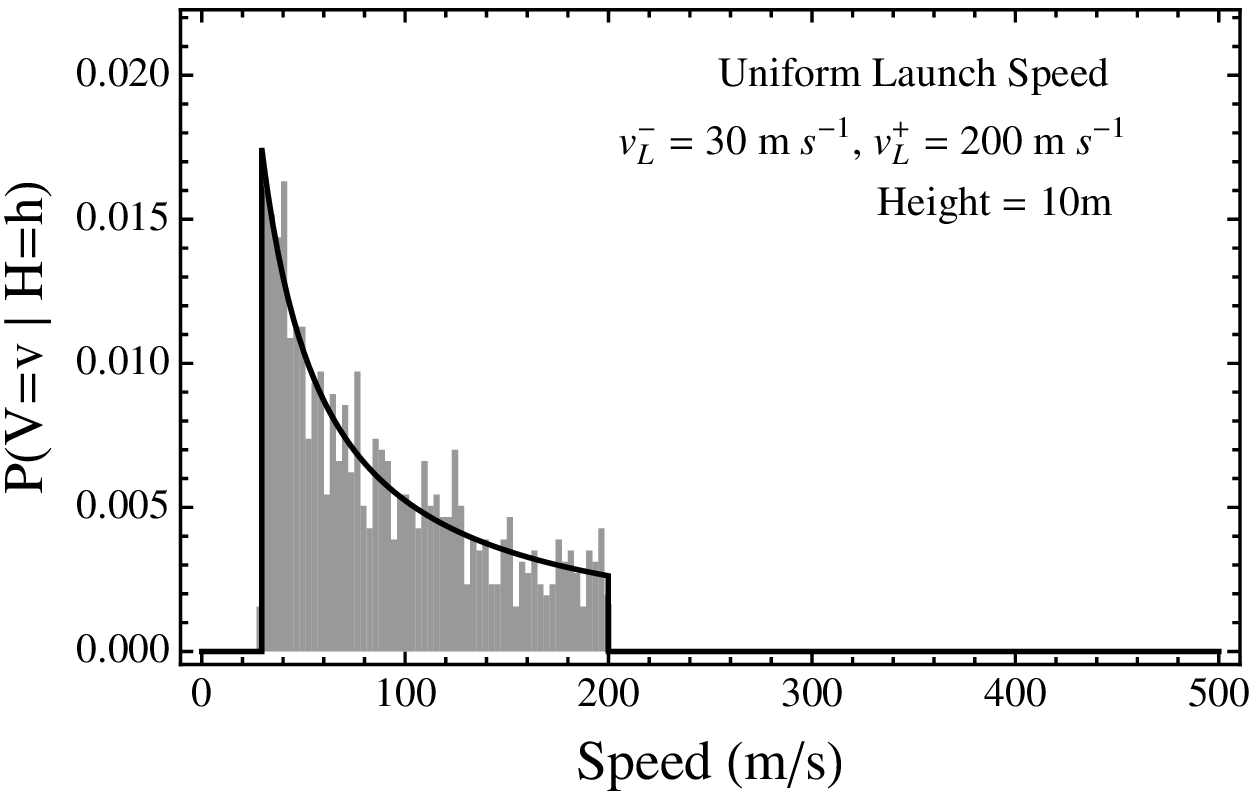}
\includegraphics[height=4cm,angle=00]{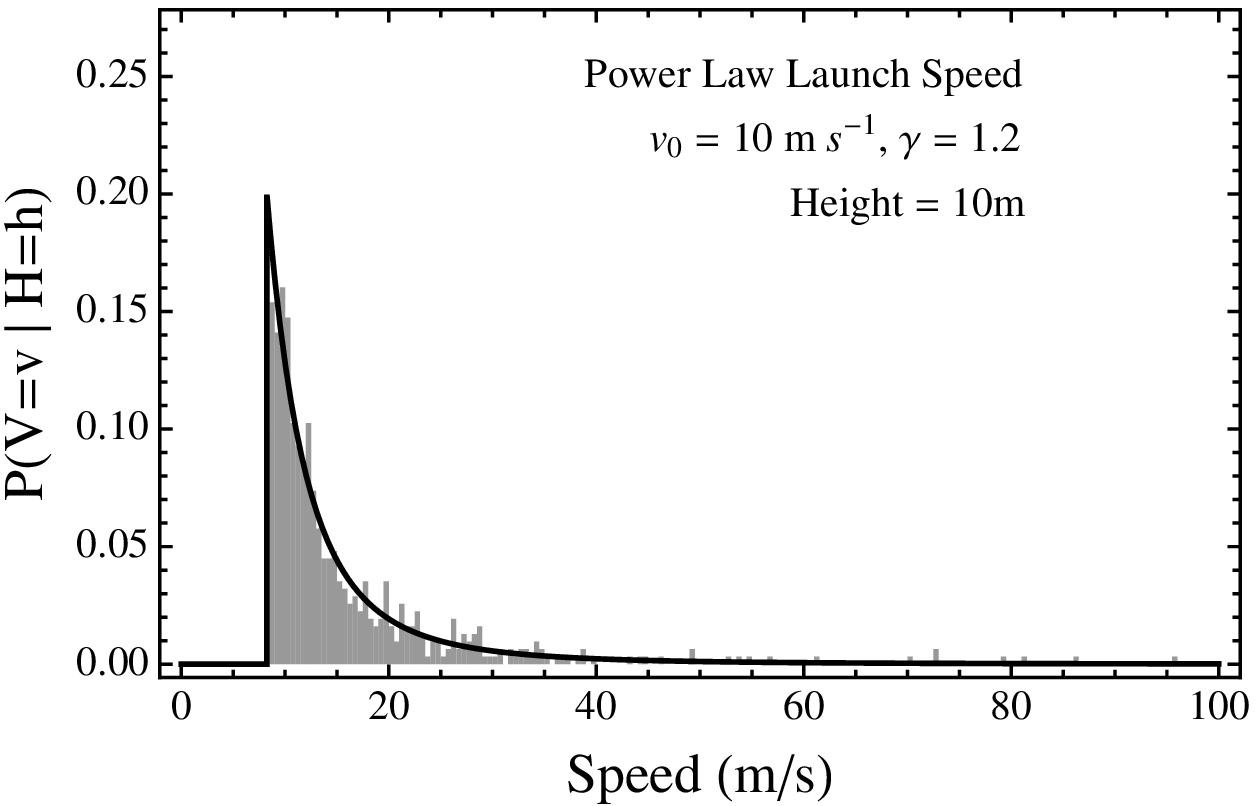}\\
\includegraphics[height=4cm,angle=00]{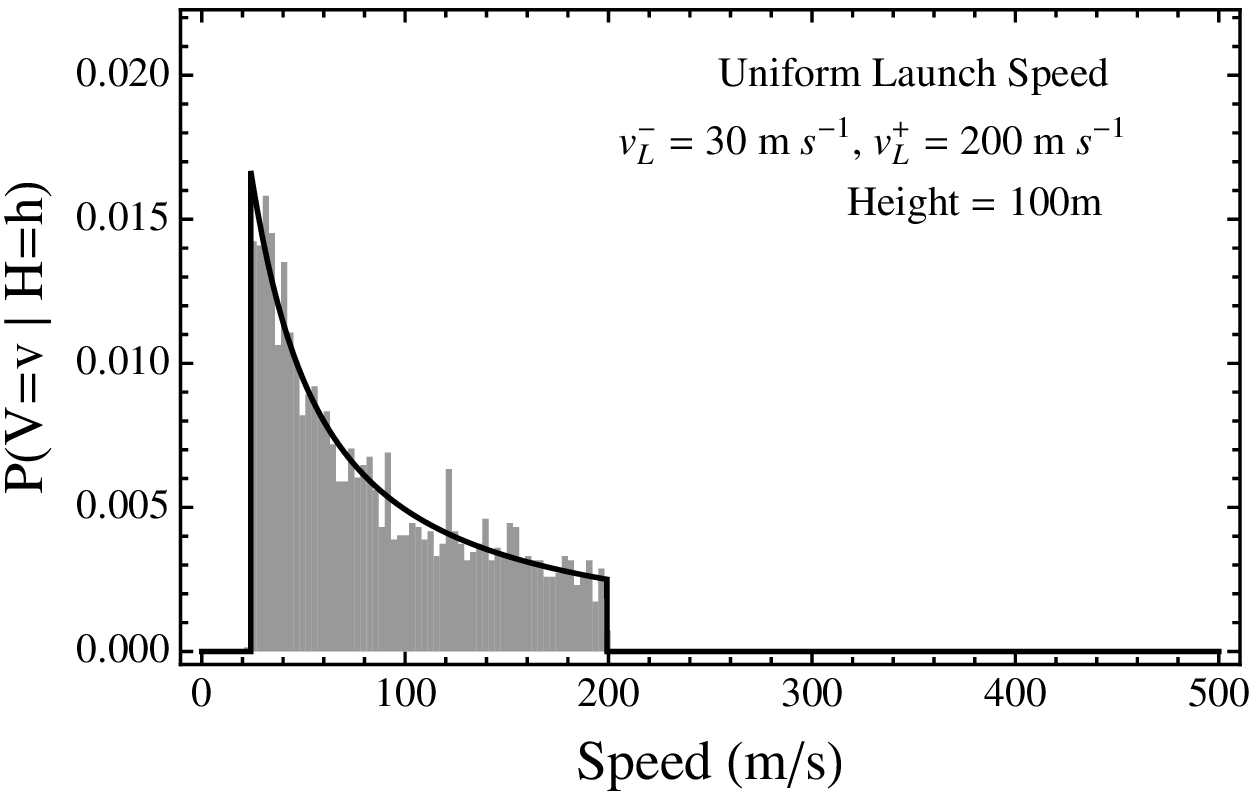}
\includegraphics[height=4cm,angle=00]{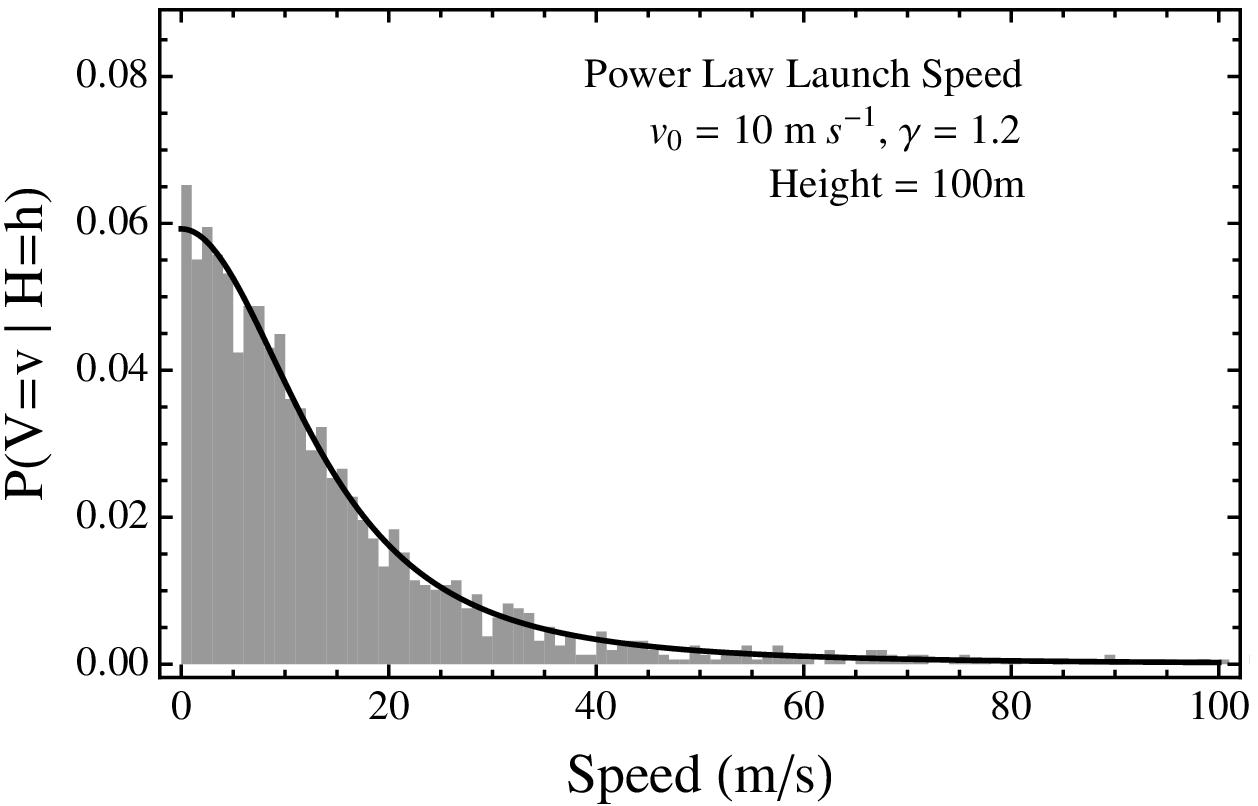}\\
\includegraphics[height=4cm,angle=00]{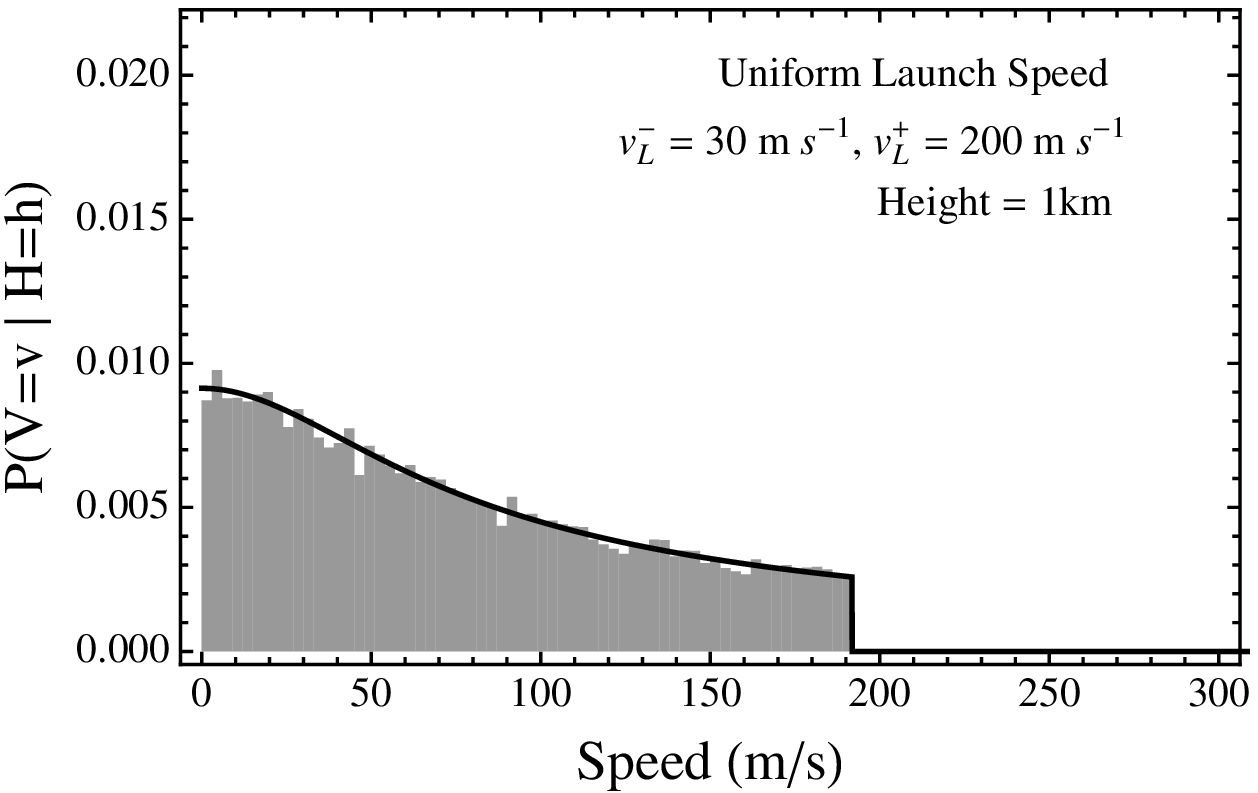}
\includegraphics[height=4cm,angle=00]{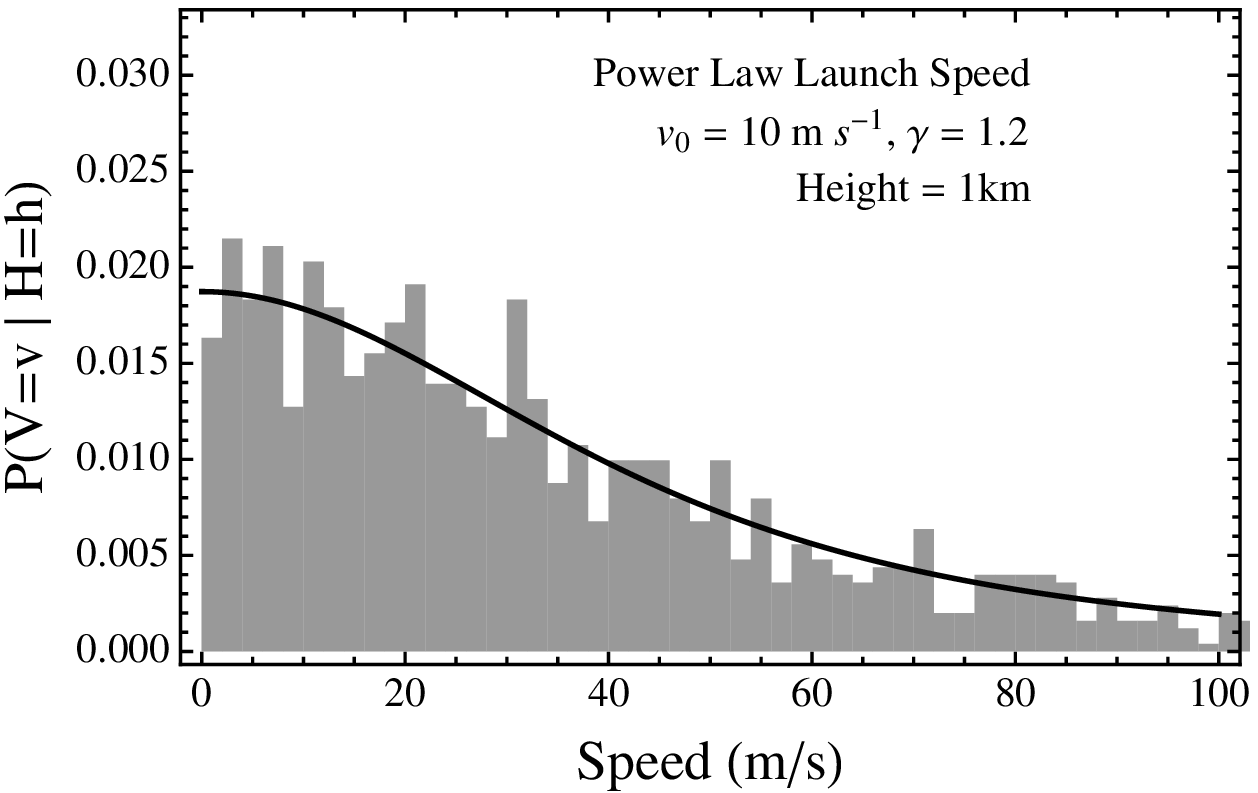}\\
\includegraphics[height=4cm,angle=00]{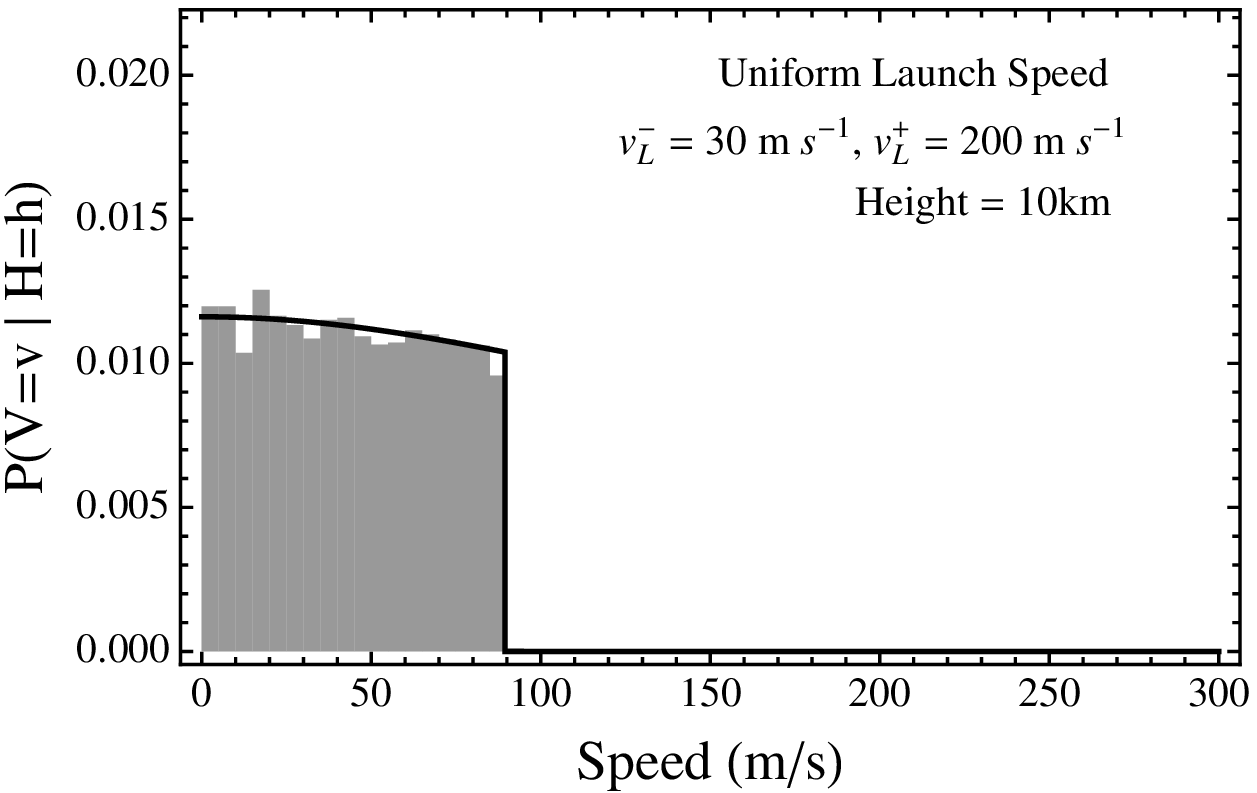}
\includegraphics[height=4cm,angle=00]{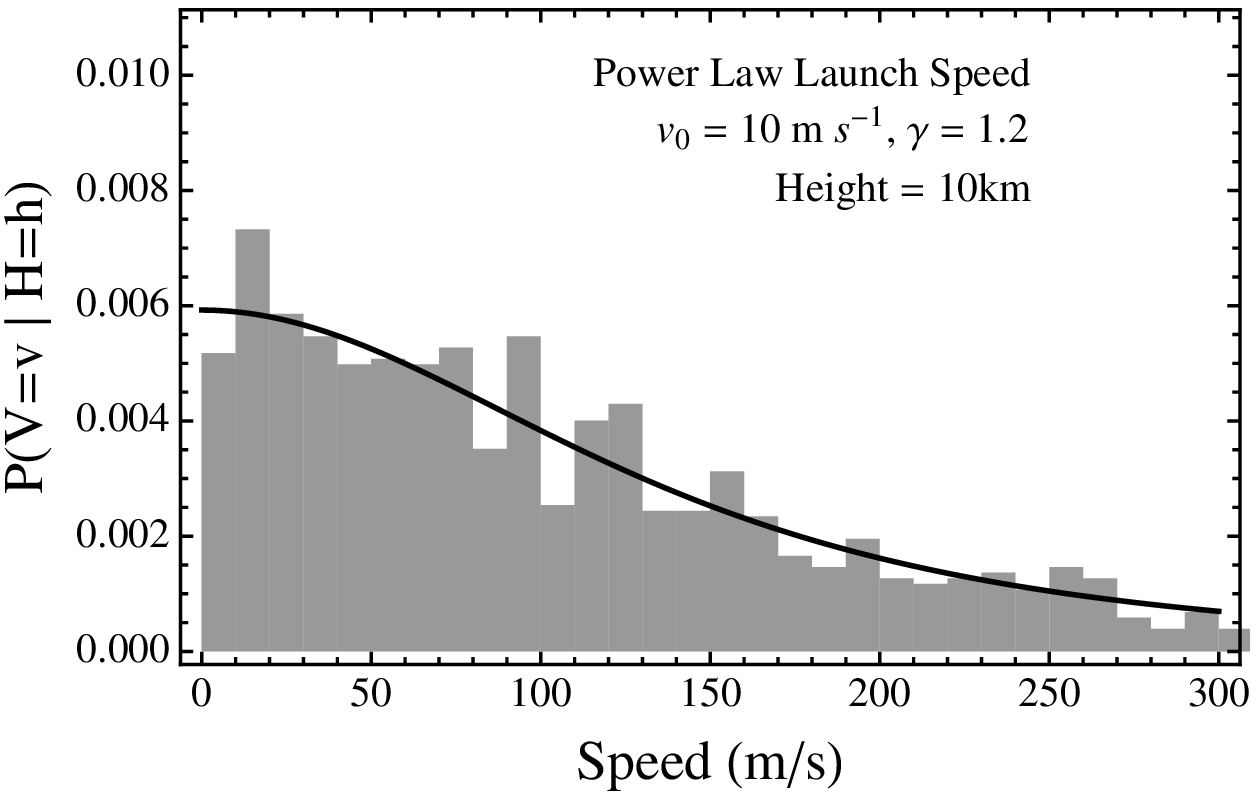}
\caption[Model (bold curves) and simulated (gray histogram) distributions of ${\rm v}$ given $h$ for a uniformly distributed (left panels) and power-law distributed (right panels) ejection speeds.]{}
\label{fig:v_given_h_distr}
\end{figure}

\begin{figure}
\vspace{-3cm}
\centering
\includegraphics[height=8cm,angle=00]{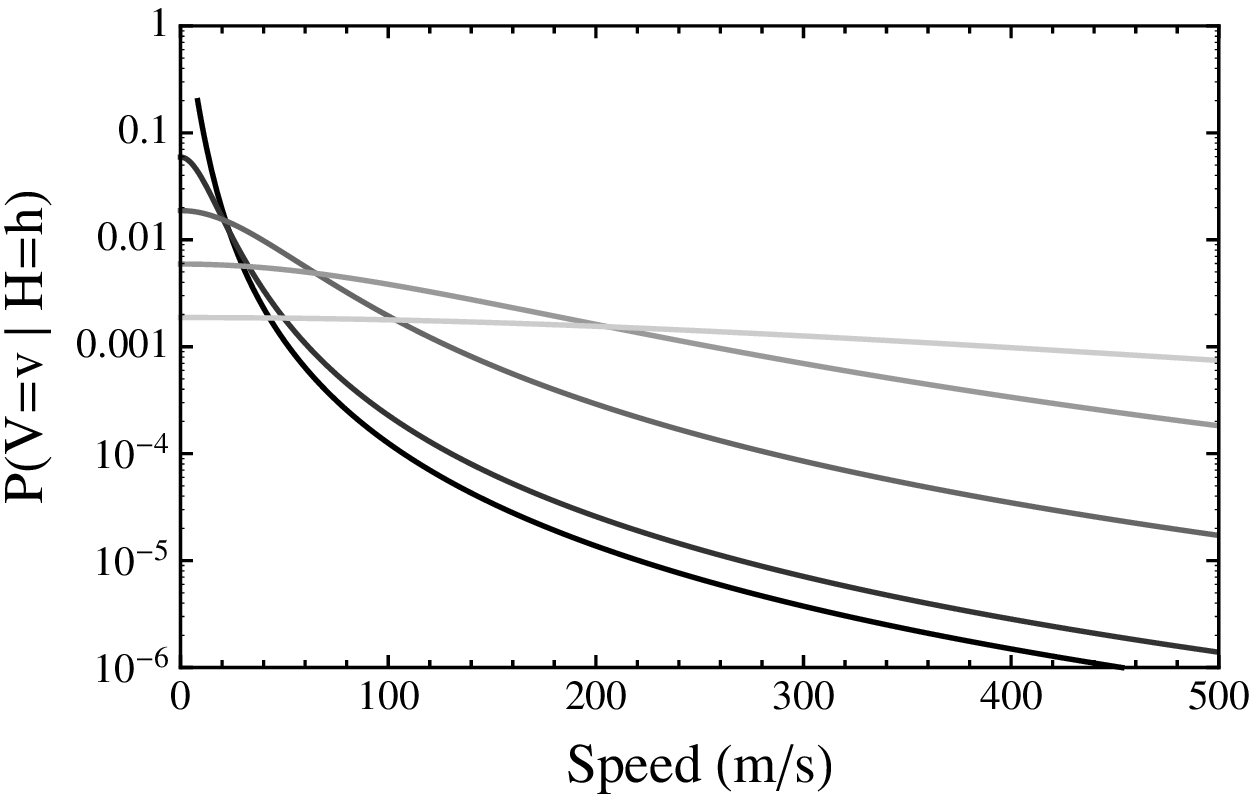}\\
\includegraphics[height=8.2cm,angle=00]{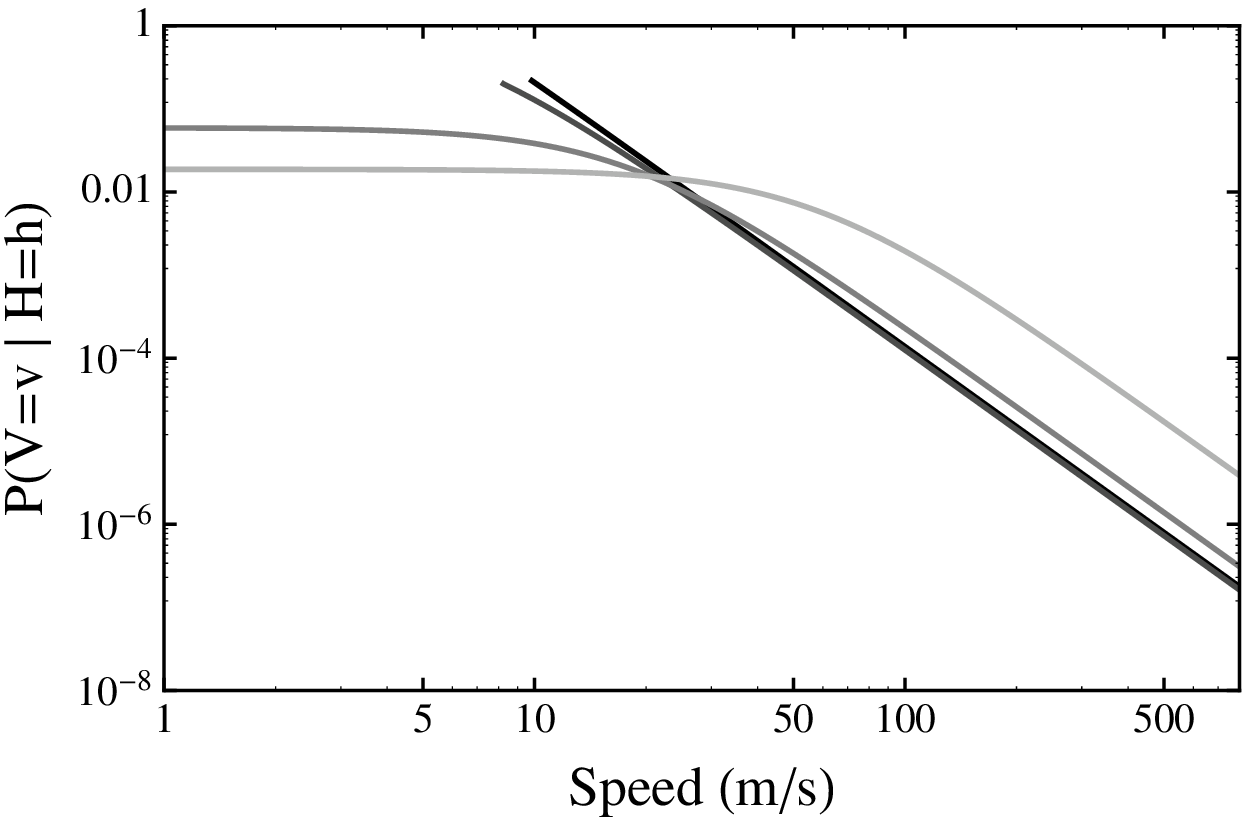}
\caption[Distribution of ${\rm v}$ given $h$ for a power-law distribution of ejection speeds for
five different values of the altitude $h$ where the speed scale is linear (top) and for four different values of the altitude on a logarithmic speed scale (bottom). A lighter colour corresponds to a higher altitude.]{}
\label{fig:v_given_h_alt}
\end{figure}

 In Fig.~\ref{fig:v_given_h_distr} we show the analytical expressions for the uniform (left panels) and the power law (right panels) distributions
 compared with speed statistics of $N = 10^{7}$ particles at altitudes - from top to bottom -  of 10 m, 100 m, 1000 m and 10 km. The parameters of the uniform distribution were set
 at ${\rm v}^{-} = 30$ m $\mbox{sec}^{-1}$ and ${\rm v}^{+} = 200$ m $\mbox{sec}^{-1}$; for the power-law distribution, $\gamma$ was set to $1.2$ and ${\rm v}_{0}$ to 10 m $\mbox{sec}^{-1}$.  Two features are worthy of note: (i) the distributions at lower altitudes are truncated at some non-zero value of {\rm v} due to the existence of a minimum value for ${\rm v}_{L}$, and (ii) the power under the distributions  for the power-law ejection speed gradually spreads to a wider range of ${\rm v}$ values, with the result that particles with speeds in excess of 100 m $\mbox{sec}^{-1}$ are unlikely for $h\leq1000$m but as likely as those with speeds below that value for $h=10$km. In any case, most of the particles would be concentrated at the lower altitudes, as the middle right panel of Fig.~\ref{fig:hv_distr} indicates. The spreading of the power towards higher speeds with increasing altitude may be seen more clearly in the top panel of Fig.~\ref{fig:v_given_h_alt} where we have plotted Eq.~\ref{eq:v_given_h_gamma} for altitudes starting from 10 m above the surface (black curve) and increasing, by factor-of-10 increments, to $10^{5}$ m (100 km; light grey curve). In particular, particle speeds $\geq 100$ m $\mbox{sec}^{-1}$ are an order of magnitude more probable at $1000$ m than at $10$ and $100$ m while at $100$ km the existence of particles with any speed up to $500$ m $\mbox{sec}^{-1}$ is comparably probable.
 
In the bottom panel of the same Figure one sees the same expression plotted as above for $h=$1 m up to $h=$1000 m but on a logarithmic speed scale. This shows that, at the lowest altitudes (e.g. $h=1$m), the conditional probability density of speed given the altitude is approximately a power-law function of speed. At higher altitudes, it is insensitive to speed up to ${\rm v} \simeq \sqrt{2gh}$, becoming a power law at higher speeds. This follows from Eq.~\ref{eq:v_given_h_gamma} where, if ${\rm v}^{2} \ll 2 g h$ then $P(V={\rm v}|H=h)$ is approximately equal to $\left(2 g h\right)^{-1/2}$ while, when the situation is reversed, $P(V={\rm v}|H=h) \propto {\rm v}^{-\left(\gamma+2\right)}$. Therefore, as found for the altitude pdf examined in Section~\ref{altitude}, the distribution of particle speeds depends on the ejection physics with the dependence being more apparent near the surface.
  
In relation to the {\it in situ} characterisation of ejecta clouds, the frame in which measurements are collected is not always inertial. It is instructive, therefore, to consider the pdf of the speed ${\rm w}$ of grains relative to a platform moving horizontally with speed $u$ (Fig.~\ref{fig:3d_ejection_geometry}). In this case, we have 
 \begin{equation}  \label{eq:w_given_h}
P(W={\rm w}|H=h) =P(V={\rm v}\left(u\mbox{, }{\rm w}\right) | H=h) \mbox{ }{\rm w}/{\rm v}\left(u\mbox{, }{\rm w}\right)
\end{equation}

where ${\rm v}\left(u\mbox{, }{\rm w}\right)=\sqrt{{\rm w}^{2} - u^{2}}$.

\begin{figure}
\vspace{-3cm}
\centering
\includegraphics[height=4cm,angle=00]{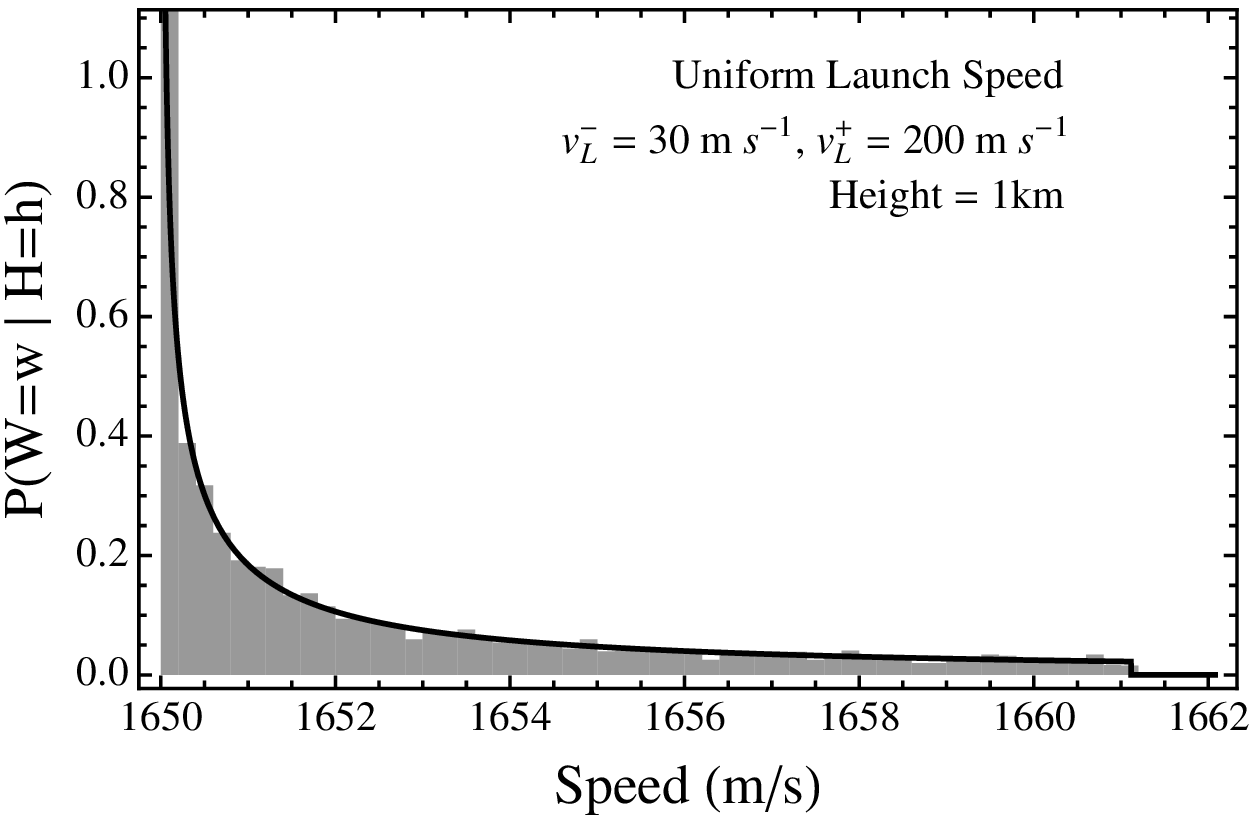}\hspace{0.5cm}\includegraphics[height=4cm,angle=00]{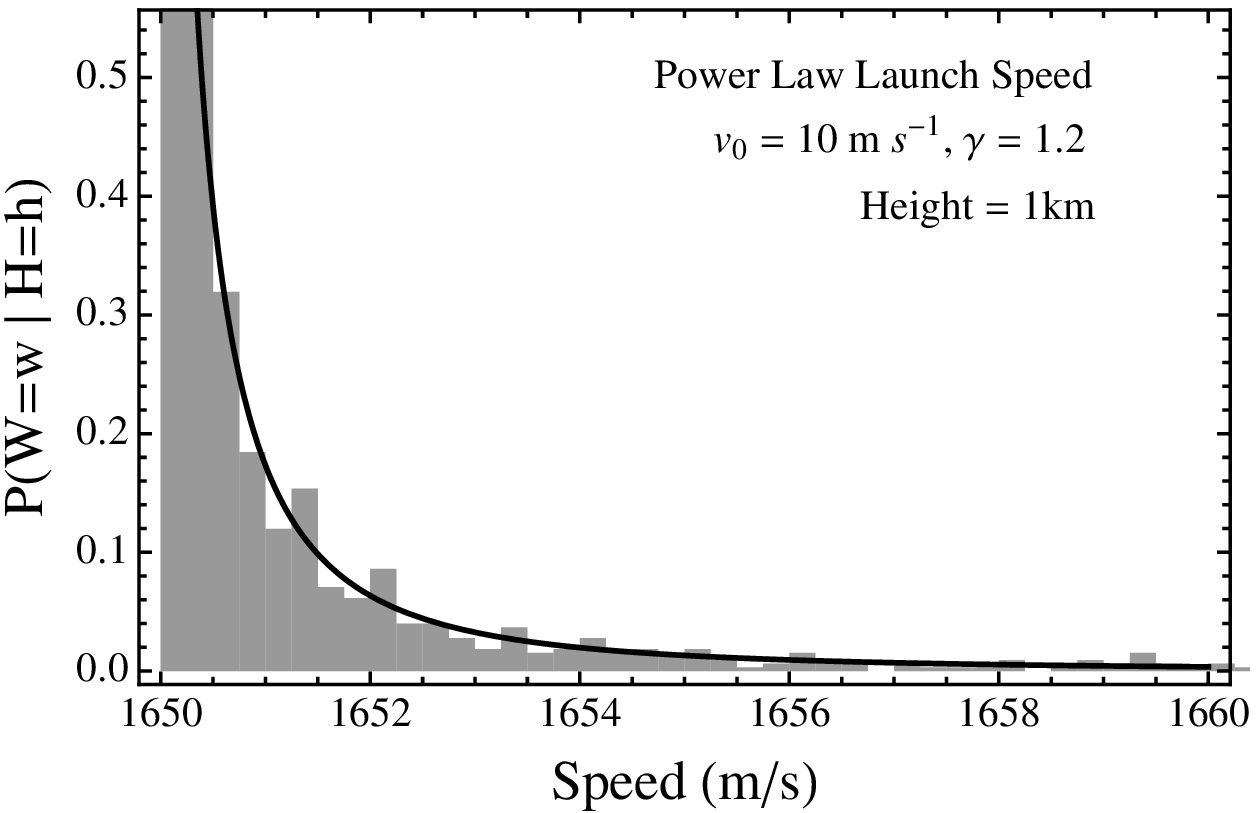}\\
\includegraphics[height=4cm,angle=00]{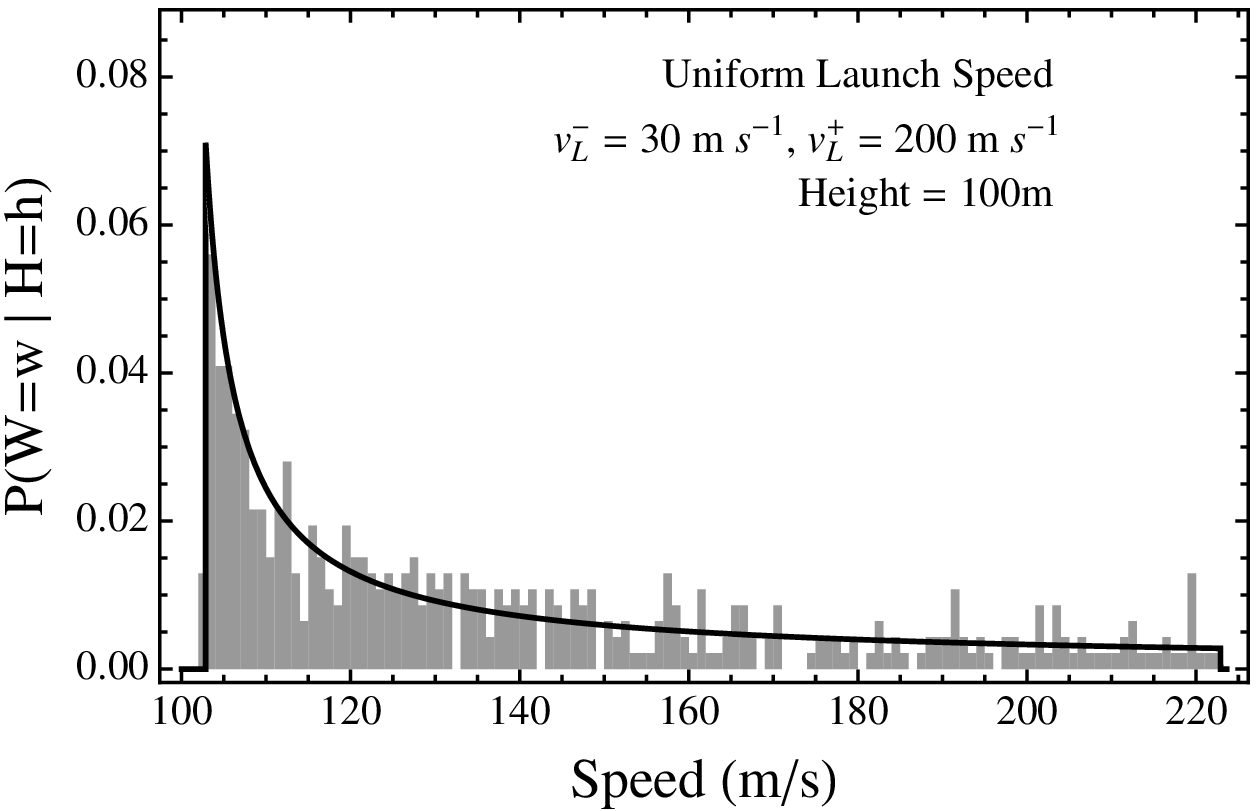}\hspace{0.5cm}\includegraphics[height=4cm,angle=00]{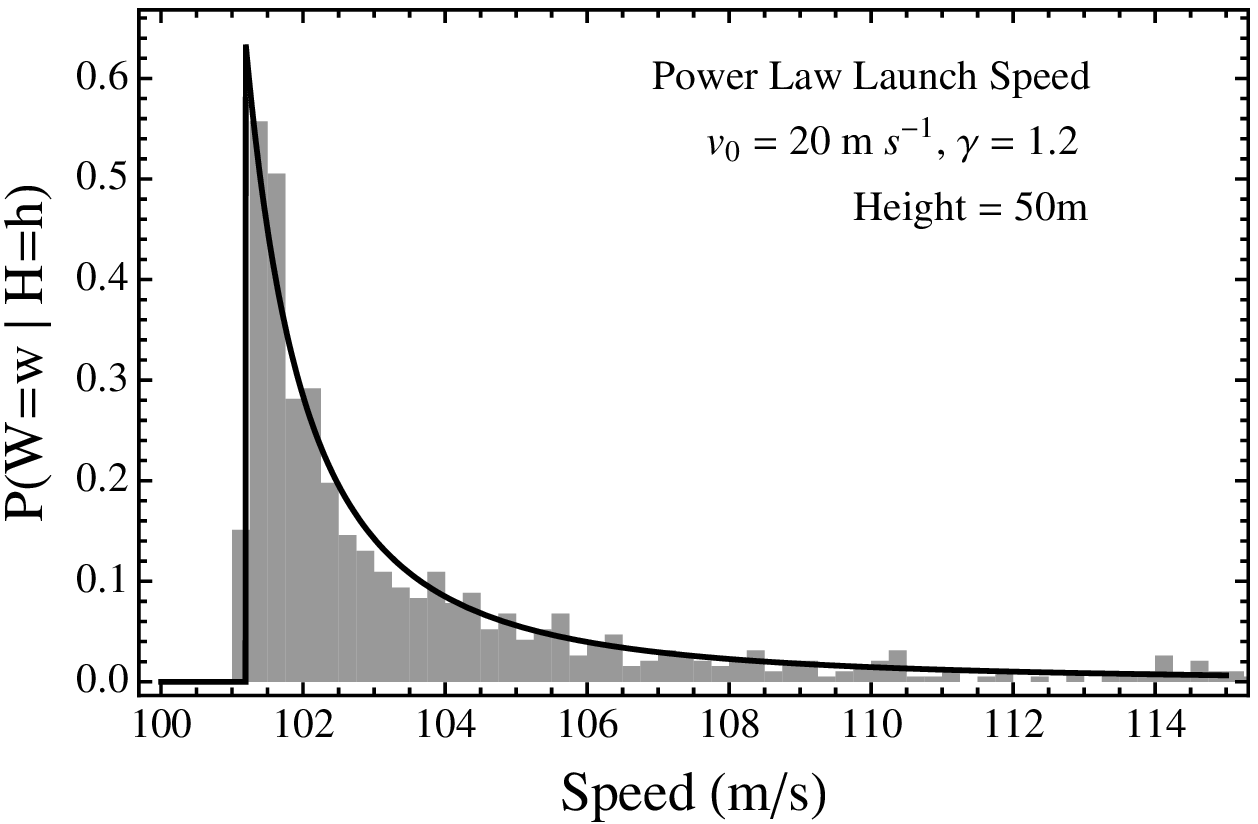}
\caption[Distribution of the speed of ejecta ${\rm w}$ at altitude $h$ relative to a moving platform for a uniform (left panels) and a power-law (right panels) distribution of ejection speeds.]{}
\label{fig:w_given_h_alt}
\end{figure}

In Fig.~\ref{fig:w_given_h_alt} we compare the analytical expression (\ref{eq:w_given_h}) with the results of numerical simulations 
and the following choices of parameter values: uniformly-distributed ejection velocities at $h=1000$ m and a platform
moving with $u=1650$ m $\mbox{sec}^{-1}$ (top left); at $h=100$ m and $u=100$ m $\mbox{sec}^{-1}$ (bottom left);
power-law-distributed ejection velocities at $h=1000$ m, $u=1650$ m $\mbox{sec}^{-1}$ and ${\rm v}_{0}=10$ m 
$\mbox{sec}^{-1}$ (top right); and at $h=50$ m, $u=100$ m $\mbox{sec}^{-1}$ and ${\rm v}_{0}=20$ m $\mbox{sec}^{-1}$ (bottom right).
The simulations with uniformly-distributed ejection speeds were run with $N=10^{6}$ particles while those for power-law-distributed ejection speeds
utilised $N=10^{7}$ particles.

\section{Three-dimensional particle motion}
\label{3d_motion}
\begin{figure}
\vspace{-3cm}
\centering
\includegraphics[width=21cm,angle=90]{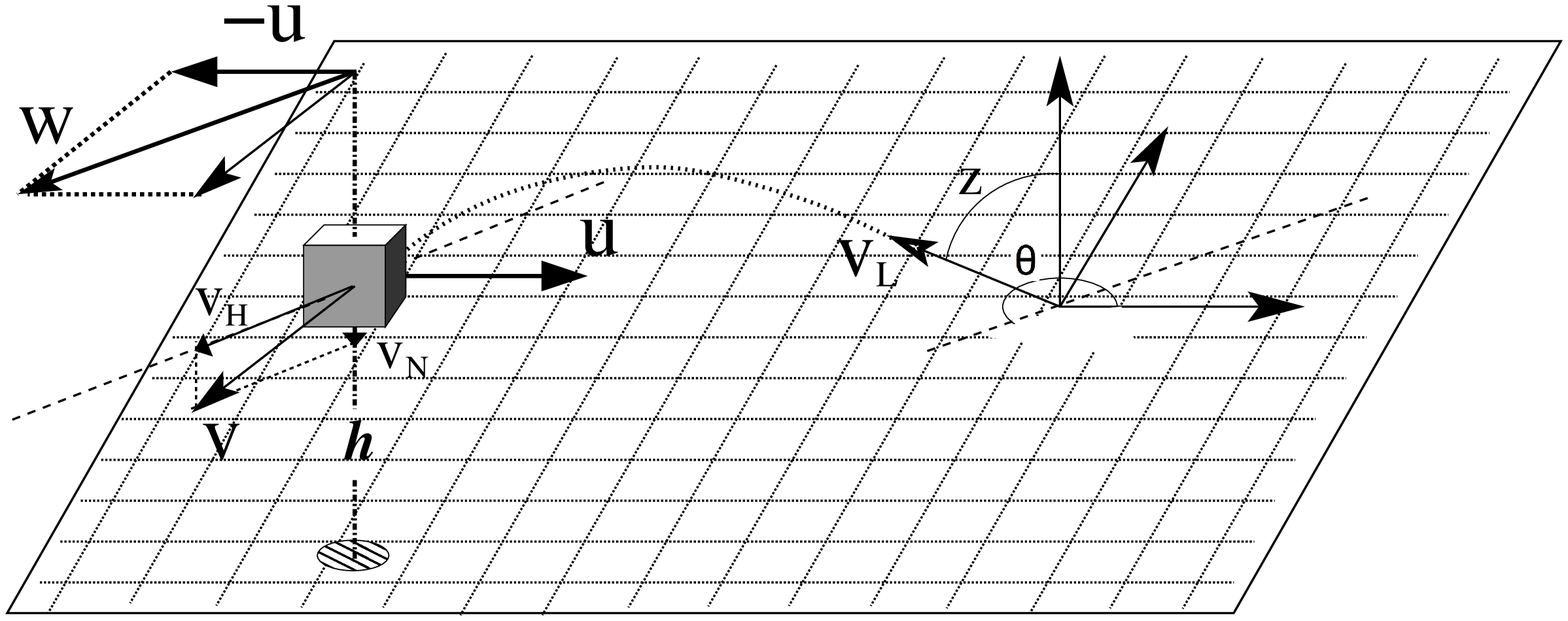}
\caption[Graphical representation of the kinematics of an ejected particle relative to the surface and to a moving platform as implemented in our model. See main text for notation.]{}
\label{fig:3d_ejection_geometry}
\end{figure}

 If grain motion is allowed in all three dimensions, the ejection velocity vector ${\rm v}_{L}$ need not be vertical but at an angle $z$ to the surface normal (Fig.~\ref{fig:3d_ejection_geometry}). The (constant) horizontal component of the velocity ${\rm v}_{H}={\rm v}_{L} \sin z$ propagates the particle at an azimuth $\theta$ to the (arbitrary) reference direction. We adopt the notation ${\rm v}_{N}$ for the grain velocity component normal to the surface to distinguish it from ${\rm v}_{L}$. Laboratory experiments to-date provide evidence for preferential ejection of material at a narrow range of zenith angles \citep{Gault.et.al1963,KoschnyGrun2001b} \cite[see also discussion in][]{Kruger.et.al2000} so we adopt here a single value of $z$ as a simplifying yet physically realistic assumption.  

Under this new notation, the ejection speed laws described by Eqs.~\ref{eq:pdf_v} and \ref{eq:pdf_v_gamma} remain formally correct. The ensemble altitude and speed distributions derived in the previous Sections incorporate the pdf for what is now ${\rm v}_{N}$. The expressions given for the altitude (Section~\ref{altitude}) remain valid if the parameters ${\rm v}_{0}$, ${\rm v}^{+}$ and ${\rm v}^{+}$ are replaced with their respective projections onto the surface normal, that is ${\rm v}_{0N}={\rm v}_{0} \cos z$, ${\rm v}^{+}_{N}={\rm v}^{+} \cos z $ and ${\rm v}^{-}_{N}={\rm v}^{-} \cos z$. In what follows, we provide speed distributions for the case of power-law distributed ejection velocities (Eq.~\ref{eq:pdf_v_gamma}).
 
To derive the pdf of the speed ${\rm v}$ given the altitude $h$, one considers that there is now an additional random variable, the launch azimuth $\theta$ (see Fig.~\ref{fig:3d_ejection_geometry}), uniformly distributed in the interval $[0\mbox{, }2\pi\hspace{-0.05cm})$ and independent of ${\rm v}_{L}$ and $\Delta t_{L}$. One can use this fact to write the joint probability density of
$\theta$, ${\rm v}$ and $h$ as
  \begin{eqnarray}\label{eq:pdf_th_vl_dtl} 
P  (\Theta = \theta(\theta\mbox{, }h\mbox{, }{\rm v}) \mbox{, }V_{L}={\rm v}_{L}(\theta\mbox{, }h\mbox{, }{\rm v})  \mbox{, }\Delta T_{L}=\Delta t_{L}(\theta\mbox{, }h\mbox{, }{\rm v}))   ||J|| d \theta d{\rm v} dh  & = &  \nonumber \\
  \frac{1}{2 \pi} P(V_{L}={\rm v}_{L}  \mbox{, }\Delta T_{L}=\Delta t_{L}) d \theta d {\Delta t_{L}} d {\rm v}_{L}  &  &
 \end{eqnarray} 
where $J$ is the Jacobian of the transformation. It can be shown that the joint pdf of ${\rm v}_{L}$ and $\Delta t_{L}$ is equal to the pdf for the 1-D case divided by $\cos z$. The determinant of the Jacobian is equal to $\left({\rm v}/{\rm v}_{N}\right) {\rm v}^{-1}_{L}$. Note that this expression evaluates to ${\rm v}^{-1}_{L}$ for $z=0^{\circ}$ ie one recovers the result for vertical ejection as intuitively expected. To arrive at the desired pdf, one integrates with respect to $\theta$ and makes use of Eq.~\ref{eq:pdf:v_given_h}. Since the integrand is independent of $\theta$, the integration is trivial and the result is
 \begin{equation}\label{eq:v_given_h_3d}
P(V = {\rm v} | H = h) = \frac{g (\gamma -1)}{2 \cos z} {\rm v}^{\gamma -1}_{0} {\rm v}^{ - \gamma - 2}_{L}(h\mbox{, }{\rm v}) \frac{{\rm v}} {{\rm v}_{N}(h\mbox{, }{\rm v})} \frac{1}{f_{P}(h\mbox{; }{\rm v}_{0N})}
\end{equation}
where ${\rm v}^2_{L}={\rm v}^{2} + 2 g h$,  ${\rm v}^2_{N}={\rm v}^{2}\cos^{2} z - (2 g h) \sin^{2} z$ and $f_{P}$ is given by Eq.~\ref{eq:pdf_h_for_p}.

\begin{figure}
\vspace{-3cm}
\centering
\includegraphics[height=4cm,angle=00]{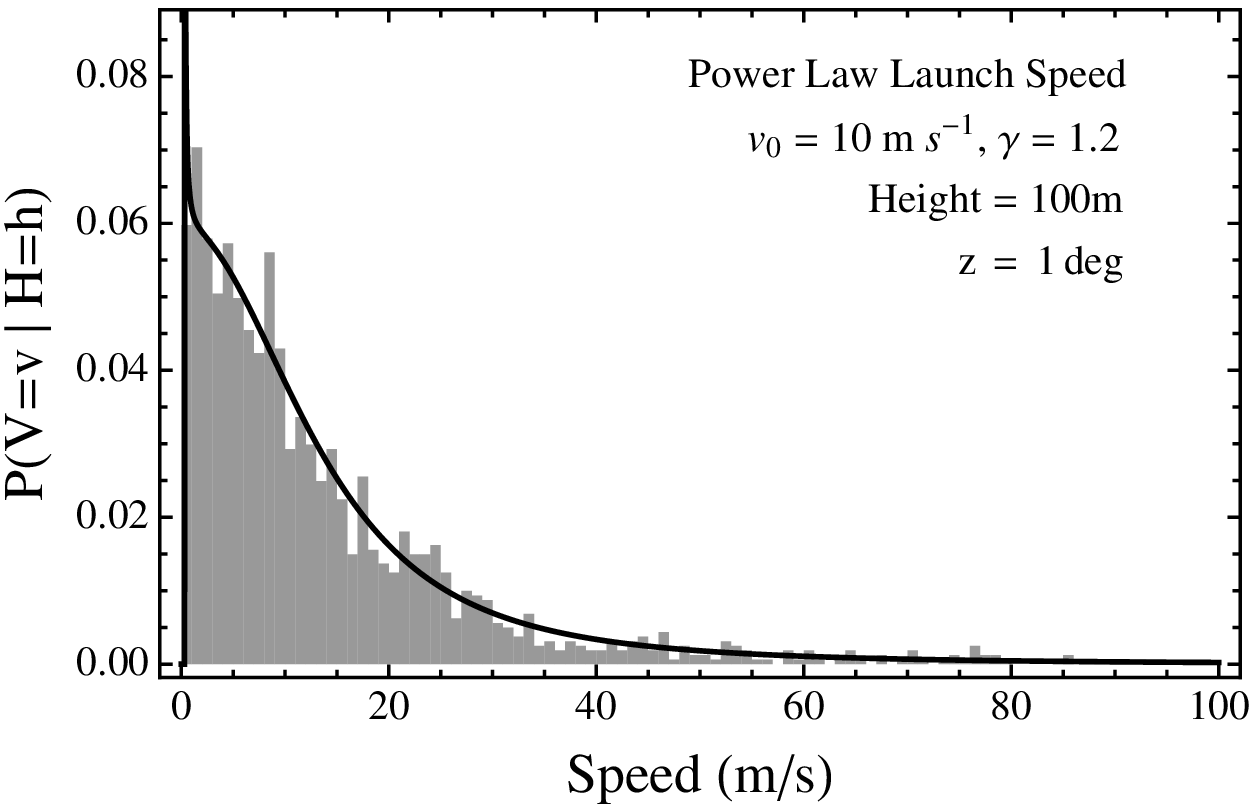}\hspace{0.5cm}\includegraphics[height=4cm,angle=00]{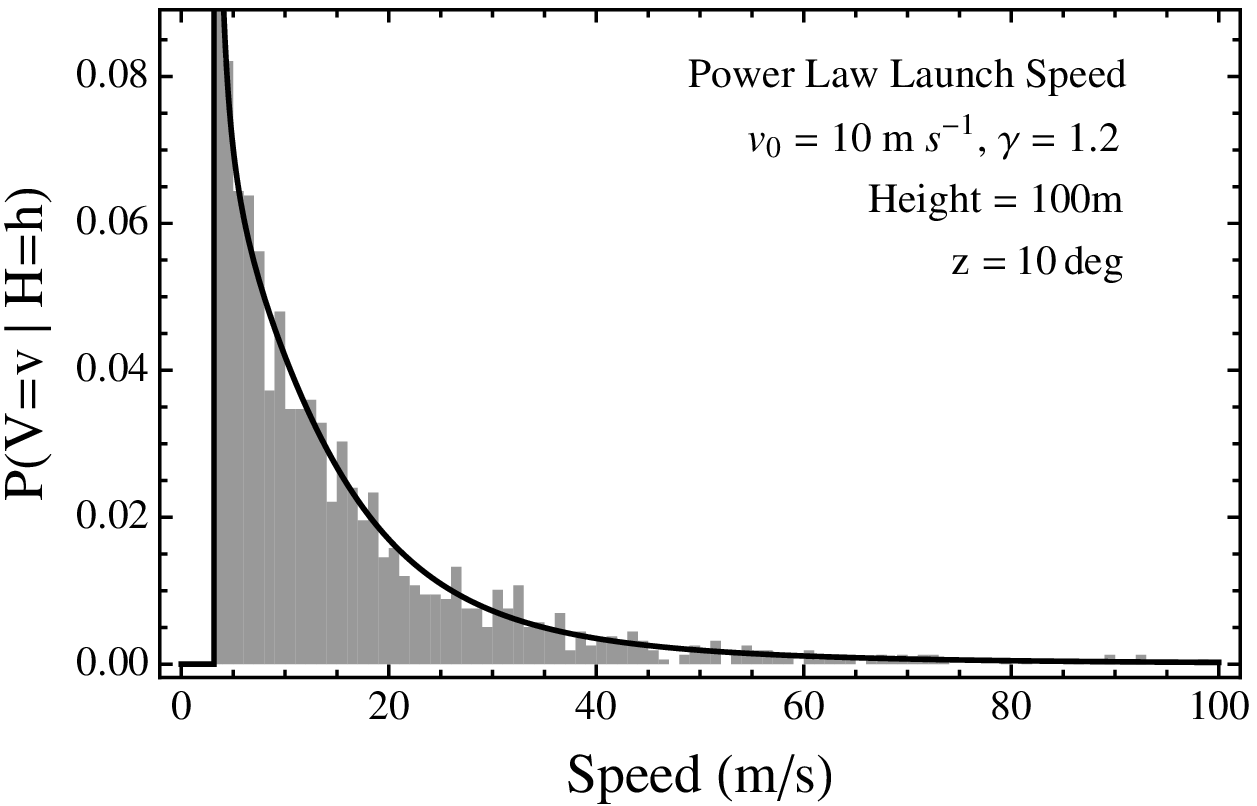}\\
\includegraphics[height=4cm,angle=00]{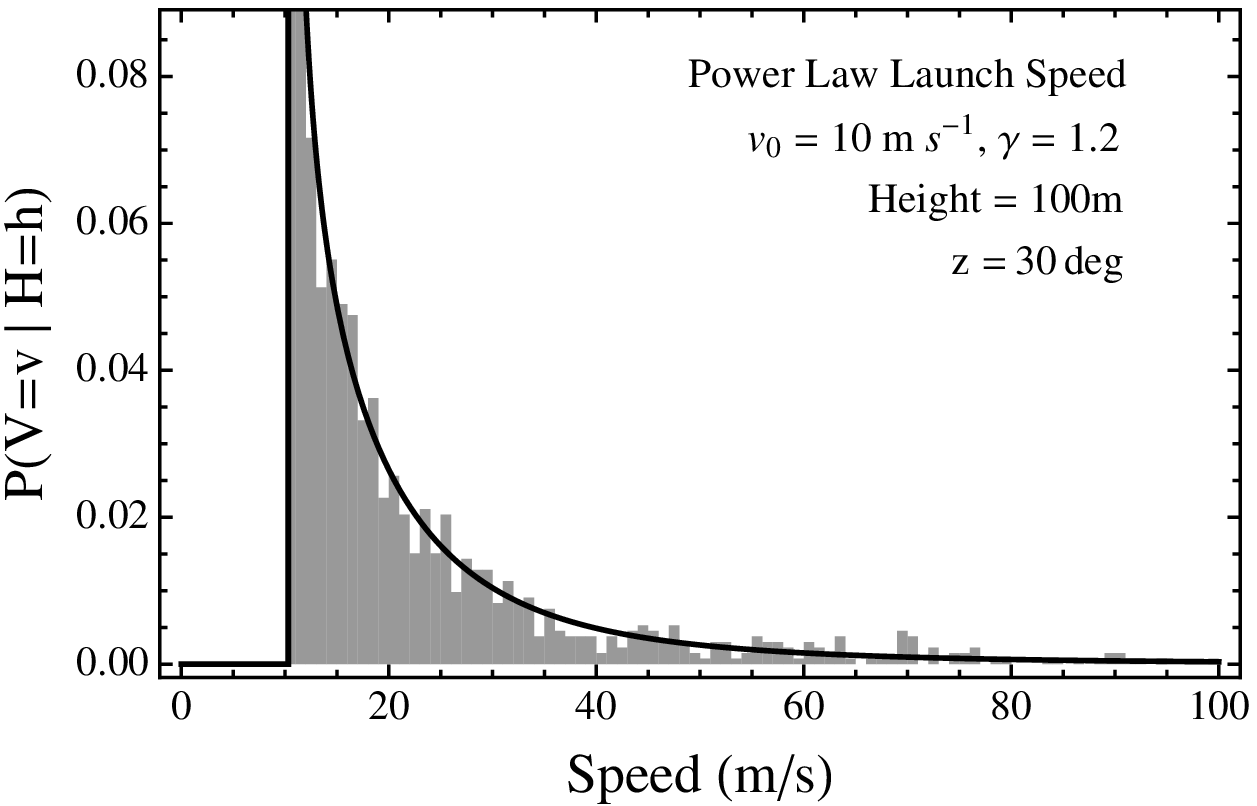}\hspace{0.5cm}\includegraphics[height=4cm,angle=00]{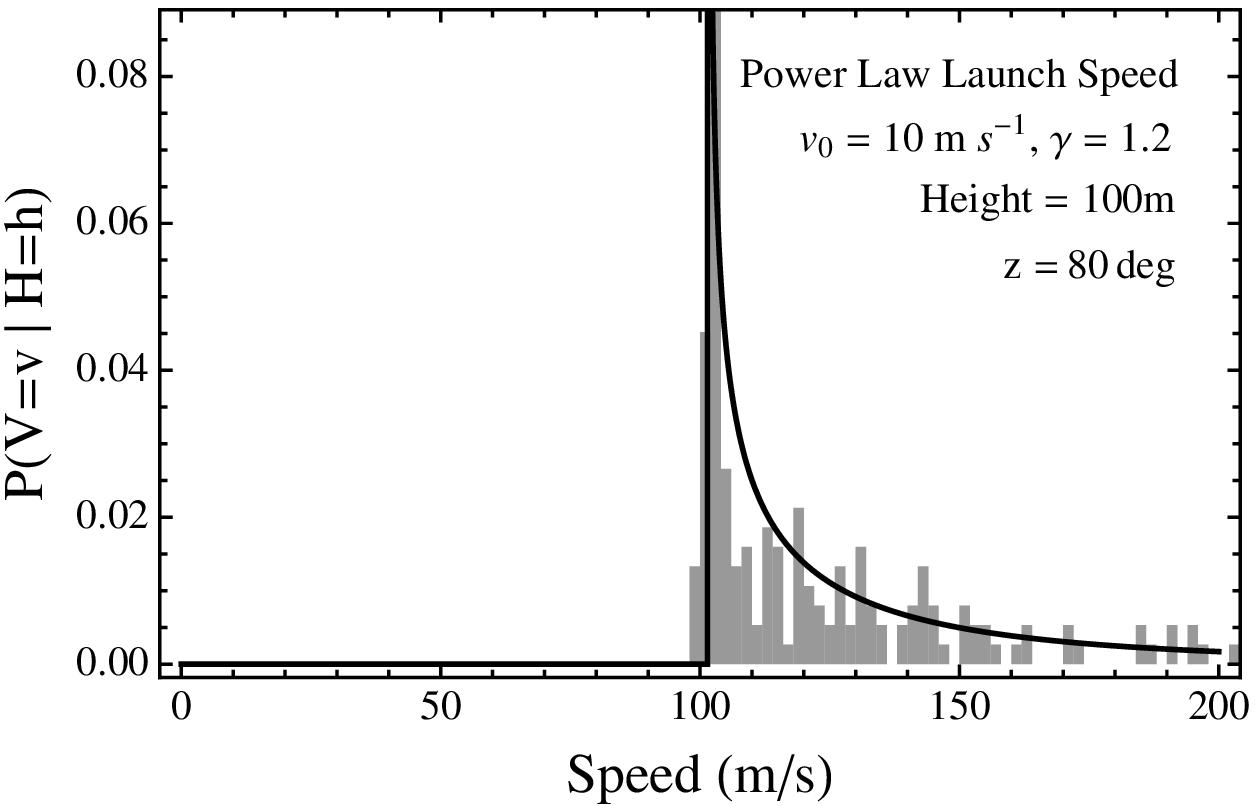}
\caption[Distribution of the speed ${\rm v}$ of ejecta at altitude $h$ in the three-dimensional problem for a power-law distribution of ejection speeds and for different values of the zenith angle $z$ of ejection.]{}
\label{fig:v_given_h_3d}
\end{figure}

Fig.~\ref{fig:v_given_h_3d} shows this pdf at an altitude of $100$m superimposed on the results of numerical simulations as $z$ increases from $1^{o}$ (top left) to $10^{o}$ (top right), $30^{o}$ (bottom left) and finally $80^{o}$ (bottom right). The adopted values of the model parameters are ${\rm v}_{0} = 10$ m $\mbox{sec}^{-1}$, $\gamma=1.2$. For reference, the case of vertical ejection for the same parameter values is represented by the second plot down from the top right of Fig.~\ref{fig:v_given_h_distr}. The cutoff at low speeds is due to the constraint that ${\rm v}_{N} \geq 0$ and corresponds to a mode at ${\rm v} = \tan z  \sqrt{2 g h}$. As a general comment on the speed distributions in this paper, it is tempting - in the absence of knowledge to the contrary and if experimentally-observed counts are relatively low as is the case in Fig.~\ref{fig:v_given_h_3d} - to intuitively expect the speed statistics to follow a different distribution eg.~a maxwellian. This, however, would {\it not} be correct, at least under the assumptions of our model. In the final part of the paper, the potential implications of lifting some of those assumptions are discussed.
 
For a platform moving horizontally at speed $u$, $\theta$ can be measured from the direction of motion as Fig.~\ref{fig:3d_ejection_geometry} shows.
In this reference frame, the speed ${\rm w}$ of the particle relative to the platform may be written as
 \begin{equation} \label{eq:w_vh_theta}
{\rm w}^{2} =  u^{2} + {\rm v}^{2} - 2 u \sqrt{2 g h +{\rm v}^{2}} \sin z \cos \theta \mbox{.}
 \end{equation}
   
 To derive the probability of ${\rm w}$ given $h$, we consider the transformation $\left( \theta\mbox{, }{\rm v}\right) \rightarrow  \left( \theta\mbox{, }{\rm w}\right)$ for which
\begin{equation} \label{eq:jacobian_v_w}
 ||J|| = \frac{{\rm w}}{{\rm v}} \left(1 - u \sin z \cos \theta /  \sqrt{2 g h +{\rm v}^{2}} \right)^{-1}
\end{equation}  
   
The sought-for pdf is then  
 \begin{equation}\label{eq:w_given_h_3d}
P(W={\rm w}|H=h) = \int_{0}^{2 \pi} \frac{P(V={\rm v}\left({\rm w}\mbox{, }\theta\right)\mbox{, }\Theta=\theta\mbox{, }H=h)}{f_{P}(h\mbox{; }{\rm v}_{0N})} ||J|| d \theta
\end{equation}
where the notation used in the integrand is from Eqs.~\ref{eq:pdf_th_vl_dtl} and \ref{eq:v_given_h_3d}.

Note that it is necessary to solve Eq.~\ref{eq:w_vh_theta} for ${\rm v}$. This is a bi-quadratic which, in the first instance, admits to the real roots
\begin{eqnarray}\label{eq:solns_v1v2}
{\rm v}^{2}_{1,2} &=&  \left[ {\rm u}^2 \left(2 \sin^{2} z \cos^{2} \theta -1 \right) + {\rm w}^2 \right] \nonumber \\
& &\pm 2 {\rm u} \sin z \cos \theta \sqrt{ {\rm w}^2 + 2 g h + {\rm u}^2 \left(\sin^{2} z \cos^{2} \theta - 1 \right)}
\end{eqnarray}
if
\begin{equation}\label{eq:v_wrt_th_w_req1}
{\rm w}^{2} \geq {\rm u}^{2}\left(1 - \sin^{2} z \cos^{2} \theta \right) - 2 g h \mbox{.}
\end{equation} 

Positive roots lead to solutions for ${\rm v}$ (hereafter referred to as ``solutions'' to distinguish them from the roots of Eq.~\ref{eq:w_vh_theta}) and define the subsets of the interval $[0\mbox{, }2 \pi]$ - as functions of ${\rm w}$ - over which the integration in $\theta$ takes place in Eq.~\ref{eq:w_given_h_3d}. If, for given values of $\theta$ and ${\rm w}$ (say, $\theta^{\ast}$ and ${\rm w}^{\ast}$) only one of the two solutions exists, the integral of Eq.~\ref{eq:w_given_h_3d} is evaluated for that solution only. If both solutions exist, the conditional probability density evaluates to $ P({\rm v} \left( {\rm w}^{\ast},\theta^{\ast}\right)\mbox{, } \theta^{\ast}\mbox{ } |\mbox{ } h) =  P({\rm v}_{1}\mbox{, }\theta^{\ast}\mbox{ }  |\mbox{ }  h) + P({\rm v}_{2}\mbox{, }  \theta^{\ast}\mbox{ }  | \mbox{ } h)$. A useful quantity for computational purposes is the product of the two roots of Eq.~\ref{eq:w_vh_theta}, equal to
\begin{equation}\label{eq:product_v_squares}
\left({\rm w}^{2} - {\rm u}^{2}\right)^{2} - 8 {\rm u}^{2} g h \sin^{2} z \cos^{2} \theta \mbox{.}
\end{equation}

\begin{figure}
\vspace{-3cm}
\centering
\includegraphics[width=12cm,angle=0]{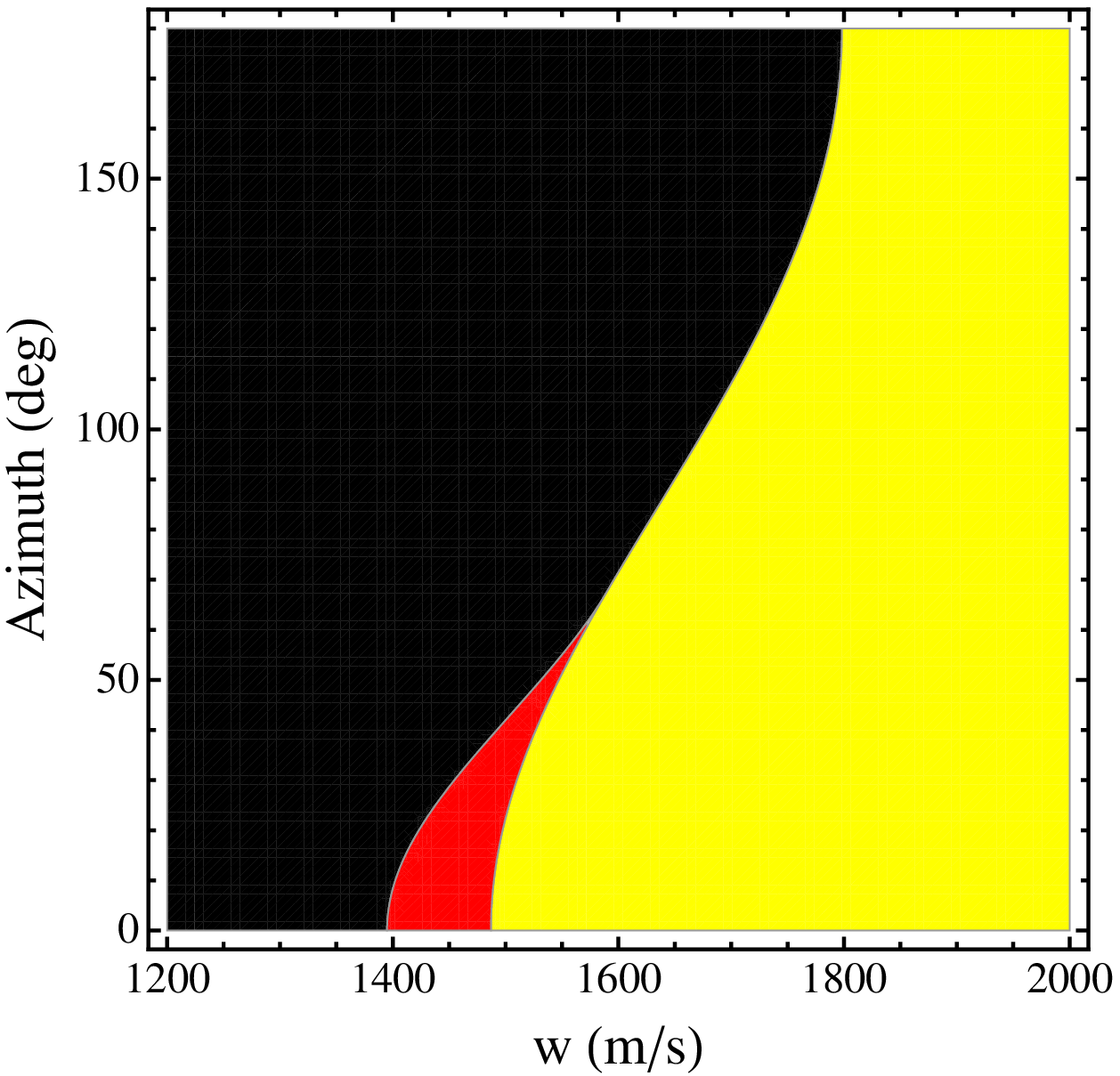}
\caption[Partition of $\left({\rm w}, \theta \right)$ space in terms of the number of solutions for the speed ${\rm v}$ of the particle in Eq.~\ref{eq:w_vh_theta} for $z=30^{\circ}$.]{}
\label{fig:w_theta_3d}
\end{figure}

To illustrate the partition of $\left({\rm w},\theta\right)$ space in terms of the existence and the number of real solutions for ${\rm v}$ in Eq.~\ref{eq:w_vh_theta} we have colour-coded the different domains accordingly in Fig.~\ref{fig:w_theta_3d} where $z = 30^{\circ}$, $u=1650$ m $\mbox{sec}^{-1}$ and $h=30$ km.  In the black region, the bi-quadratic has no real, positive solutions for ${\rm v}$ that satisfy Eq.~\ref{eq:w_vh_theta}. In the 
yellow region, exactly one such solution exists, namely ${\rm v}_{1}$ (Eq.~\ref{eq:solns_v1v2}). Finally, in the red region, both ${\rm v}_{1}$ and ${\rm v}_{2}$ are distinct, real and positive solutions of Eq.~\ref{eq:w_vh_theta}. 

\begin{figure}
\vspace{-3cm}
\centering
\includegraphics[width=14cm,angle=0]{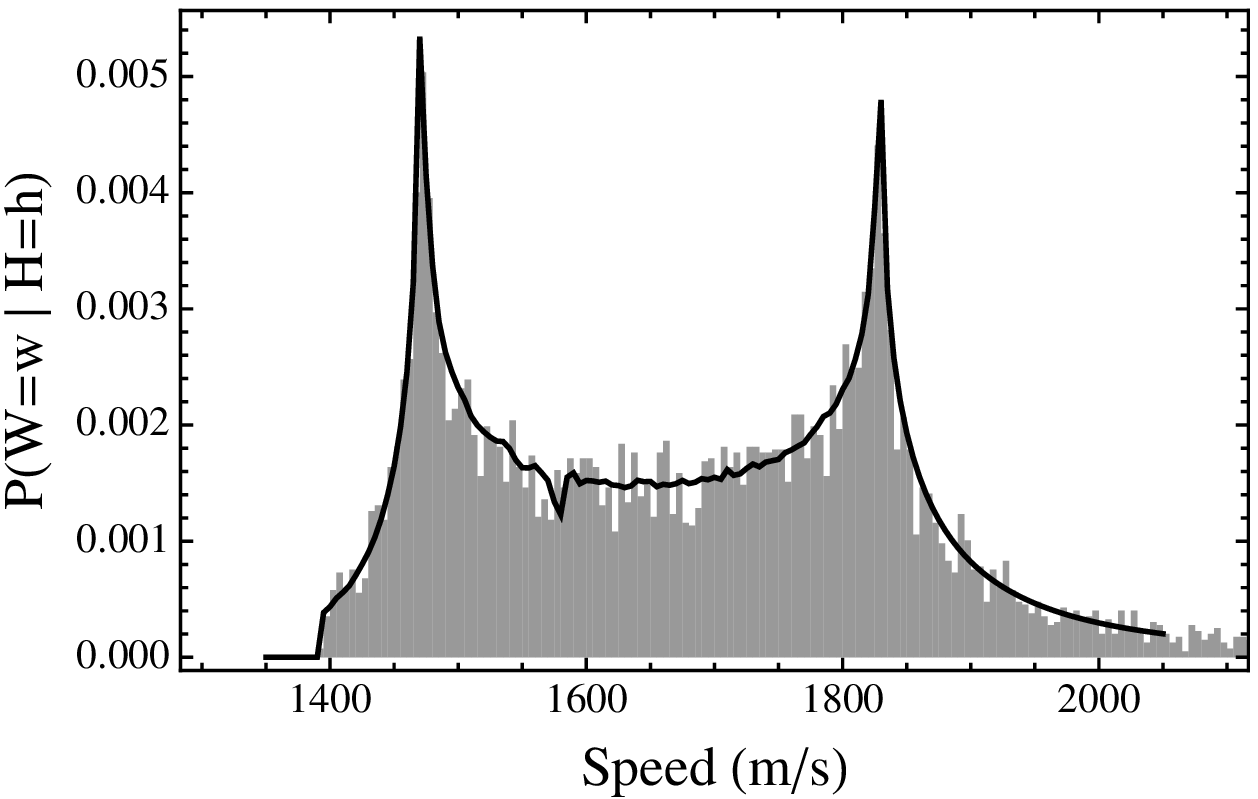}\\\includegraphics[width=14cm,angle=0]{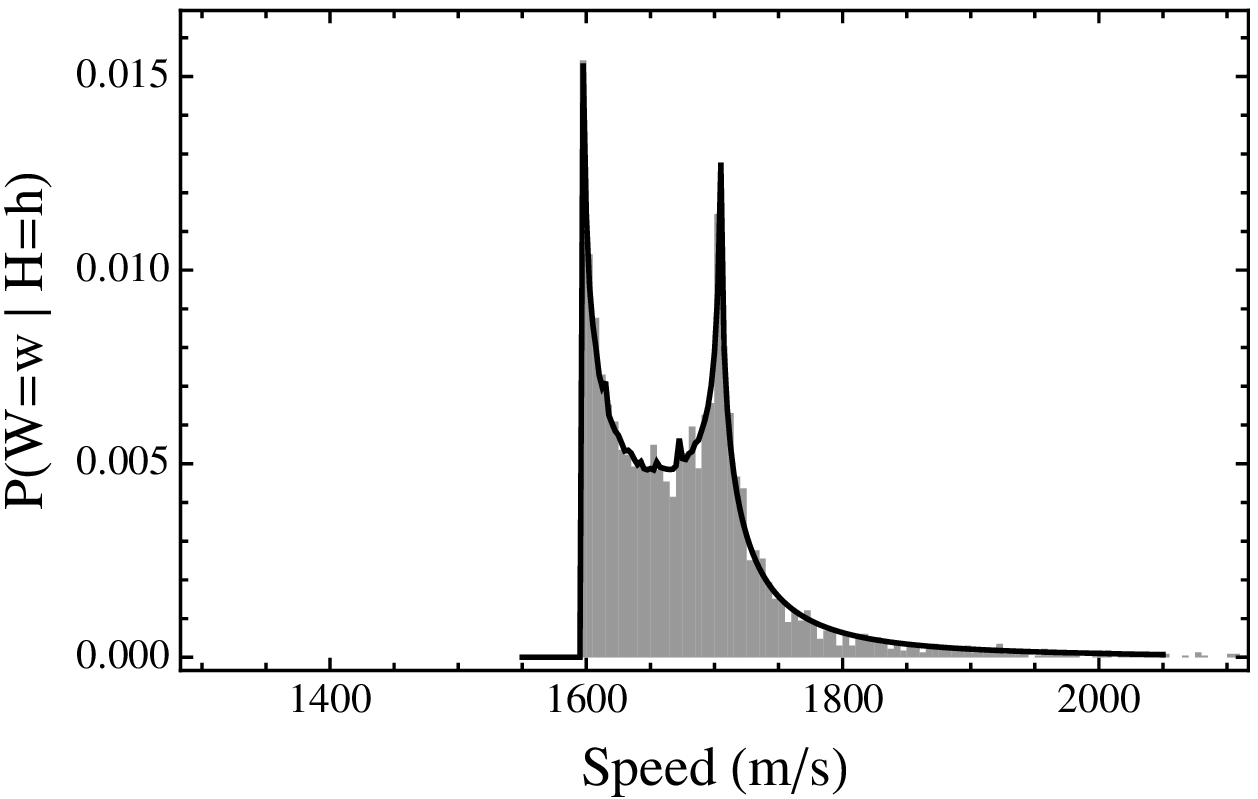}
\caption[Distribution of the speed of ejecta ${\rm w}$ relative to a platform moving horizontally for a power-law distribution of ejection speeds in the three-dimensional problem and for $z=30^{\circ}$ (top) and $z=10^{\circ}$ (bottom).]{}
\label{fig:w_3d}
\end{figure}

The model probability distribution is illustrated in the top panel of Fig.~\ref{fig:w_3d} against the statistics of $10^{7}$ Monte Carlo variates launched with velocities distributed according to Eq.~\ref{eq:pdf_v_gamma} with ${\rm v}_{0}=100$ m $\mbox{sec}^{-1}$ and $\gamma=1.2$. The integral in Eq.~\ref{eq:w_given_h_3d} has been evaluated numerically using the {\tt NIntegrate} subroutine within the {\it Mathematica} package \citep{Wolfram2010}.
The main features of the observed distribution are the cutoff in the probability density below 1400 m $\mbox{sec}^{-1}$ and the two peaks at 1470 m $\mbox{sec}^{-1}$ and 1830 m $\mbox{sec}^{-1}$ respectively. 
These are related to the boundaries between the different regions of $\left({\rm w},\theta\right)$ space identified in Fig.~\ref{fig:w_theta_3d}. By experimenting with different values of $h$, ${\rm v}_{0}$ and $\gamma$, we find that these are generic features of the distribution of the impact speed of ejecta on a moving platform in orbit and at altitudes of 10-100 km above the surface. The two peaks approach each other as $z$ decreases (compare the top and bottom panels of Fig.~\ref{fig:w_theta_3d}, the latter showing the same distribution but for $z=10^{\circ}$). They merge into one as $z$ vanishes, reverting back to the expressions for vertical ejection (Fig.~\ref{fig:w_given_h_alt}). Finally, we comment on the ``dent'' that appears in the distribution for $z=30^{\circ}$ at ${\rm w} \sim 1580$ m $\mbox{sec}^{-1}$. It is not a real feature of the distribution, but arises due to the difficulty in evaluating the integral in Eq.~\ref{eq:w_given_h_3d} over the thinning part of the red domain as ${\rm w} \rightarrow {\rm u}$ and ${\rm \theta} \rightarrow \pi/2$ (Fig.~\ref{fig:w_theta_3d}).

  \section{Size-dependent ejection speed}
In the preceding sections we have considered ejection speed distributions which do not depend on ejecta size. On the basis of past work, it is reasonable to expect a dependence of the ejection speed on the sizes of both the ejectum and the impactor \citep{Melosh1984,Miljkovic.et.al2012}. In that case, the ejection speed probability will be given by
 \begin{equation}  \label{eq:vl_s_d}
P(V_{L}={\rm v}_{L})=\int_{s_{I,min}}^{s_{I,max}} \int_{s_{E,min}}^{s_{E,max}} P(V_{L}={\rm v}_{L}  \mbox{, }S_{E}=s_{E} \mbox{, }S_{I}=s_{I}) \mbox{ }d s_{E} \mbox{ }d s_{I}
 \end{equation}   
 where $s_{E}$ and $s_{I}$ denote the sizes (diameters) of the ejectum and impactor respectively. The integrand can be expressed
 in terms of the probability distributions of these two quantities through the chain rule for conditional probability:
  \begin{eqnarray}  \label{eq:vl_chain_rule}
P(V_{L}={\rm v}_{L}\mbox{, }S_{E}=s_{E} \mbox{, }S_{I}=s_{I}) &=&  P(V_{L}={\rm v}_{L}|S_{E}=s_{E} \mbox{, }S_{I}=s_{I}) \nonumber  \\
             &  & P(S_{E}=s_{E}|S_{I}=s_{I}) P(S_{I}=s_{I})\mbox{.}
 \end{eqnarray}   
 To demonstrate the use of these expressions, we provide an example below for particular choices of the different distributions. For the impactor size range of interest, the last term on the right-hand-side can be approximated by a power law \citep{Grun.et.al1985}
  \begin{equation} \label{eq:pdf_s_imp}
 P(S_{I}=s_{I}) \approx \alpha \left( s_{I}/s_{I,min} \right)^{-\alpha}/s_{I} \mbox{,    }s_{I,min} < s_{I}<s_{I,max}
 \end{equation} 
 if $s^{-\alpha}_{I,min} \gg s^{-\alpha}_{I,max}$.  
 In the recent study of the ejecta clouds of the jovian moons Europa and Ganymede by \citet{Miljkovic.et.al2012} the distribution of ejecta sizes was taken to be a deterministic function of ejection speed. In a probabilistic framework, this corresponds to the
 degenerate pdf for the first term on the right-hand side of Eq.~\ref{eq:vl_chain_rule}:
   \begin{equation}\label{eq:pdf_vl_given_si_sf}\hspace{-0.25cm}
 P(V_{L}={\rm v}_{L}|S_{E}=s_{E} \mbox{, }S_{I}=s_{I})=1 \mbox{ if }{\rm v}_{L}=C \left( s_{E} /s_{I} \right)^{-k}\mbox{ and }0\mbox{ otherwise}
 \end{equation}
 where the constant $C$ depends on the target surface properties. For the ejecta size distribution, we adopt a power law
    \begin{equation}\label{eq:pdf_sf_given_si}
      P(S_{E}=s_{E}| S_{I}=s_{I}) = \beta \left(s_{E}/s_{E,min}\right)^{-\beta} / s_{E} \mbox{,    }s_{E,min} < s_{E}<s_{E,max}
      \end{equation}
      where, for simplicity, we have assumed independence on impactor size. As with (\ref{eq:pdf_s_imp}), we require that  $s^{-\beta}_{E,min} \gg s^{-\beta}_{E,max}$.
The exponents $\alpha$, $k$ and $\beta$ are all assumed to be positive. Upon integrating Eq.~\ref{eq:vl_s_d} we find that
      \begin{equation}\label{eq:vl_beta_k}
      P(V_{L}={\rm v}_{L}) \propto {{\rm v}_{L}}^{\left(\beta+1\right)/k}
      \end{equation}
      where the constant of proportionality (say $K$) is
            \begin{equation}
     K \approx \frac{\alpha \beta}{\alpha+\beta+1} s^{\beta}_{E,{\rm min}} s^{-\left(\beta+1\right)}_{I,{\rm min}} C^{-\left(\beta+1\right)/k}\mbox{.}
      \end{equation}
      if $s^{-\left(\alpha+\beta+1\right)}_{I,min} \gg s^{-\left(\alpha+\beta+1\right)}_{I,max}$.
      
 Comparing Eq.~\ref{eq:vl_beta_k} with Eqs~\ref{eq:pdf_v_gamma} and  \ref{eq:pdf_h_for_p}, we conclude that, for this particular parametric description of the ejection process, the profiles of ejecta number density with altitude should follow a power law with an exponent of $\frac{\beta+1}{2 k}$. As this is a positive number by definition, the result predicts that the number density of ejecta will {\it increase} with altitude. Therefore, the functional form of the distributions of the ejecta kinematics appears to be sensitively-dependent on the particular
 ejection models adopted. 

 \section{Conclusions and Discussion}
 \subsection{Main Findings}
In this paper a methodology has been described for deriving explicit steady-state probability distributions of the kinematic properties of impact-generated dust in the vicinity of a planetary surface.   
This methodology has been applied to the altitude and speed distributions of ejecta and validated against numerical simulations. Below we summarise the main findings:
 \begin{itemize}
 \item  the altitude and speed distributions of ejecta in a stationary frame and for power-law-distributed ejection speeds admit to analytical expressions that can be readily evaluated for arbitrary values of the defining parameters. 
 \item for power-law distributed ejection speeds, the number density of ejecta decreases with altitude as a power law and with an exponent that directly depends on the corresponding exponent of the ejection speed distribution. The scale height of the ejecta distribution, rather than being constant, increases with altitude. 
 \item close to the surface, the altitude distribution of ejecta exhibits a cusp that translates into a sheet containing the highest number density of ejected grains. The altitude and thickness of this sheet are functions of the parameters that describe the ejection speed law. Instantaneous grain speeds are power-law distributed with an exponent that, as is the case for the altitude distribution, directly depends on the corresponding exponent of the distribution of ejection speed.
 \item the power within the speed distribution of near-surface ejecta is concentrated at low - but non-zero - speeds. For vertical ejection, this distribution becomes flatter as altitude increases, eventually allowing the possibility of particles with vanishing speed. If ejection occurs at an angle to the vertical, the shape of the distribution remains qualitatively the same but a zero ejecta speed is no longer possible at any altitude.
  \item for vertical particle ejection, the distribution of ejecta speed relative to a moving platform is qualitatively similar to that for the inertial case. If ejection occurs at an angle to the vertical, the distribution becomes qualitatively different with two separate maxima that move further apart as the zenith angle of ejection increases.
 \end{itemize}
 
Although probabilistic in nature, the collection of models in this paper encapsulates most of the information necessary to predict absolute quantities such as the particle number density $n(h)$ and the flux $F(h)$. A conceptual-level algorithm to arrive at these quantities is as follows: $n(h)$ is the product of the altitude pdf (Section~\ref{altitude}) with the number of particles in flight at a given time (assumed constant for a steady state process; Eq.~\ref{eq:nhgt0}). 
For a power-law speed distribution this is given by Eq.~\ref{eq:pdf_h_for_p_denom} where the rate parameter $\lambda$ (number of particles per unit time) can be estimated via quantitative models
of ejecta production for a given impactor flux \cite[e.g.~][]{Krivov1994}. The particle flux $F(h)$ - although dependent on the orientation of the incident surface \citep{McDonnell.et.al2001} - 
may be evaluated as $\int{ n({\rm v},h)}{\rm v} d {\rm v}$. The first term in the integrand represents the number of particles with speed {\rm v} {\it and} at height $h$;
it can be expressed as $n(h) p({\rm v}|h)$ where $p({\rm v}|h)$ is the conditional pdf of the speed given the height  from Section~\ref{speed_at_altitude}. 

\subsection{Implications for {\sl in situ} measurements}
The above can be viewed as predictions for specific features to be sought in the measured properties of impact-ejected dust near airless planetary bodies. At the same time, the class of models introduced here are, by their very nature, suitable for treating planetary dust exospheres as natural laboratories of the fundamental processes of dust mobilisation and transport in the solar system. For example, ejection speed law parameters as inferred from measurements can be compared to the results of laboratory experiments and hydrocode simulations. The knowledge gained can be applied to the low-gravity regime relevant to surface processes on NEOs and small bodies in general. In addition, although the models have been constructed through the frequentist approach to probability theory, they can also be utilised by bayesian inference techniques to extract information such as the most likely launch locations and speeds of grains at the surface. This would be particularly useful in exploring the dependence between the properties of the impactors and those of their ejecta (see also point on steady state assumption below).    
 
Recent dust measurements offer a suitable proving ground for our statistical model and an opportunity to pursue some of the above objectives, wholly or in part.
The Lunar Dust Experiment ({\it LDEX}) impact ionisation dust detector operated from October 2013 until April 2014 onboard the Lunar Atmosphere and Dust Environment Explorer ({\it LADEE}) spacecraft in orbit around the Moon \citep{Delory2014}. The charge collected by the instrument allows the grain mass and speed to be estimated from laboratory-derived calibration curves \citep{JamesSzalay2014}.
Our finding of a power-law dependence of the dust distribution (Eq.~\ref{eq:pdf_h_for_p}) implies that LDEX dust counts, when binned in altitude and corrected for the different residence time of the spacecraft in each altitude bin, will also follow a power law. Unlike the case far from the body however \citep{Krivov.et.al2003}, at the low altitudes where {\it LDEX} operated the exponent of this power law directly depends on the assumed model of the ejection physics, specifically the exponent $\gamma$ of the ejection speed distribution (Eq.~\ref{eq:pdf_v_gamma}). It follows that this exponent can be measured directly from the {\it LADEE} dust counts and that any process that modifies it will result in a different power-law fit to the measurements.
Furthermore, in Section~\ref{3d_motion} it was found that adoption of a single, non-zero value for the grain ejection zenith angle leads to a double-peaked profile for the distribution of grain impact speed relative to a moving platform (Eq.~\ref{eq:w_given_h_3d} and Fig.~\ref{fig:w_3d}). Therefore - and assuming that grain impact speeds can be measured with a precision of $\sim 50$ m $\mbox{sec}^{-1}$ or better - the existence of two peaks in {\it LDEX} data would provide evidence that a particular ejection angle dominates for real impacts. Measurements best suited for this purpose would be those collected near the pericentre and apocentre of {\it LADEE}'s orbit; at those locations the radial component of the velocity vector vanishes, in line with our assumption of a horizontally-moving platform. 

\subsection{Caveats}
To place our findings in the proper context it is necessary to highlight here several important assumptions that were made in the course of this study. The following is not intended {\bf as} an exhaustive list.
 
Probably the most important is that fragment ejection speed and size were assumed to be uncorrelated. Since our probability distributions concern the relative {\it number} of ejecta independently of their size, the measurements against which they will be compared will be dominated by a relatively narrow range of sizes. Nevertheless, understanding the consequences of introducing a dependence between size and ejection speed is important, not only to gauge how the results of this paper apply to the real solar system but also to allow the exploitation of additional information, such as size, momentum and kinetic energy in both existing and future datasets. A first step in this direction has been made in the penultimate Section of the paper where a method to treat the dependency between ejecta size and speed is presented. To demonstrate it, we worked through an example for a particular functional description of the ejection process. We find a power-law dependence of the ejecta number density on altitude similar to that found in Section~\ref{altitude} but with a {\it positive} exponent that is a function of the parameters describing both the size and speed distribution of ejecta. This emphasises the point made earlier that the distributions of the ejecta kinematics are sensitive to the ejection physics and bodes well for constraining the latter through measurements by orbiters at low altitudes or by landers at the surface.   
 
A full probabilistic treatment of this class of problems using the same methodology is outside the scope of the present work, but we note that it can (a) encompass
models where surface grains are mobilised by electrostatic forces \citep{Stubbs.et.al2006,Hartzell.et.al2013}, and (b) yield conditional distributions involving 
the physical properties of the dust (e.g. distribution of grain size/mass at a certain altitude or moving at a certain speed), also relevant to the measuring 
capabilities of current and future instrumentation \citep{Hirai.et.al2010,Sternovsky.et.al2010,Carpenter.et.al2012}. In particular, it should allow one to test the efficiency of 
electrostatic vs ballistic mobilisation of dust as competing mechanisms for the production of dust exospheres.

Next, the assumption of a {\it steady state} requires that the ensemble properties of the dust population within an altitude bin are time-invariant. In practice,
 the assumption holds if the variation of the measured statistical quantity over time is significantly smaller than the measurement itself. It is not immediately clear
 that this is true, since it depends on the efficiency of the source process (impact flux and number of ejected grains produced per unit time). It is, however, possible to emulate such a state by averaging measurements over time and for as long as the source process does not vary. For the lunar case, and given the non-isotropic background meteoroid flux in the 0.1-10 mm size range \citep{CampbellBrownJones2006,CampbellBrown2007}, one may expect source variations as the surface normal to a given location on the surface scans through the full range of angles with respect to the Earth's apex every month. 
 Shorter-term variations in the impactor flux are also expected, manifesting themselves in the Earth's atmosphere as meteor {\it showers} and {\it outbursts} \citep{Jenniskens1994,Jenniskens1995}. These should be taken into account for dust cloud modelling as they add to the value of {\it in situ} measurements in understanding the ejection process for different impactor populations. 
 
Finally, high-speed impacts by the primary ejecta population \citep{Zook.et.al1984, Grun.et.al1985} will produce {\it secondary} ejecta which have not been taken into account here. If important, their low kinetic energy relative to that of primary ejecta renders them more likely to modify the properties of the ejecta cloud at the low end of the range of altitudes considered here. The probabilistic framework in which the present model has been developed should allow treatment of multiple generations of ejecta. It would be interesting to determine, for example, if the near-surface features found in Section~\ref{altitude} persist under these circumstances. This will be explored in future work.
 
\section*{Acknowledgements}
Astronomical research at the Armagh Observatory is funded by the Northern Ireland 
Department of Culture, Arts and Leisure (DCAL). The author wishes to thank Dr Mih\'{a}ly Hor\'{a}nyi and an anonymous reviewer for their comments which improved the paper considerably; and Dr David Asher, for numerous discussions during the course of this research that prevented wasting time on less-than-productive avenues of investigation and for his comments on a draft version of the manuscript.

\bibliographystyle{elsarticle-harv}
\bibliography{yicar_11385}
\clearpage
\protect
\listoffigures

\renewcommand{\baselinestretch}{1.0}
\protect
    
\clearpage
 
\renewcommand{\baselinestretch}{2.0}

\clearpage

\clearpage

\clearpage

\clearpage

\clearpage

\clearpage

\clearpage

\clearpage

\clearpage

\clearpage
\end{document}